\documentclass[pdflatex,sn-nature,referee]{sn-jnl}

\usepackage{anyfontsize}
\usepackage{graphicx}%
\usepackage{multirow}%
\usepackage{makecell}
\usepackage{amsmath,amssymb,amsfonts}%
\usepackage{amsthm}%
\usepackage{mathrsfs}%
\usepackage{dsfont}
\usepackage[title]{appendix}%
\usepackage{xcolor}%
\usepackage{textcomp}%
\usepackage{manyfoot}%
\usepackage{booktabs}%
\usepackage{algorithm}%
\usepackage{algorithmicx}%
\usepackage{algpseudocode}%
\usepackage{listings}%
\usepackage{pgffor}%
\usepackage{float}%
\usepackage{underscore}%
\usepackage{caption}%
\usepackage{adjustbox}

\begin{document}

\title[Scaling Structure-Aware Virtual Screening to Billions of Molecules with SPRINT]{Scaling Structure-Aware Virtual Screening to Billions of Molecules with SPRINT}

\author[1]{\fnm{Andrew T.} \sur{McNutt}}
\equalcont{These authors contributed equally to this work.}

\author[2]{\fnm{Abhinav K.} \sur{Adduri}} 
\equalcont{These authors contributed equally to this work.}

\author[2]{\fnm{Caleb N.} \sur{Ellington}} 

\author[5]{\fnm{Nonso N.} \sur{Duaka}} 

\author[2,4]{\fnm{Tyler J.} \sur{Katz}} 

\author[2]{\fnm{Monica T.} \sur{Dayao}} 

\author[2,3,4]{\fnm{Eric P.} \sur{Xing}} 

\author[2]{\fnm{Hosein} \sur{Mohimani}} 

\author*[1]{\fnm{David R.} \sur{Koes}}\email{dkoes@pitt.edu}

\affil[1]{\orgdiv{Computational and Systems Biology}, \orgname{University of Pittsburgh}, \orgaddress{\city{Pittsburgh}, \state{PA}, \country{USA}}}

\affil[2]{\orgdiv{Computational Biology}, \orgname{Carnegie Mellon University}, \orgaddress{\city{Pittsburgh}, \state{PA}, \country{USA}}}

\affil[3]{\orgname{Mohamed Bin Zayed University of Artificial Intelligence}, \orgaddress{\city{Masdar City}, \state{Abu Dhabi}, \country{UAE}}}

\affil[4]{\orgname{GenBio AI}, \orgaddress{\city{Abu Dhabi}, \country{UAE}}}

\affil[5]{\orgname{Grambling State University}, \orgaddress{\city{Grambling}, \state{LA}, \country{USA}}}


\abstract{Virtual screening of small molecules against protein targets can accelerate drug discovery and development by predicting drug-target interactions (DTIs). However, structure-based methods like molecular docking are too slow to allow for broad proteome-scale screens, limiting their application in screening for off-target effects or new molecular mechanisms. Recently, vector-based methods using protein language models (PLMs) have emerged as a complementary approach that bypasses explicit 3D structure modeling. Here, we develop SPRINT, a vector-based approach for screening entire chemical libraries against whole proteomes for DTIs and novel mechanisms of action. SPRINT improves on prior work by using a self-attention based architecture and structure-aware PLMs to learn a co-embedding space for drugs and targets, enabling efficient binder prediction, search, and retrieval. SPRINT achieves competitive enrichment factors in virtual screening on LIT-PCBA, DTI classification benchmarks, and binding affinity prediction benchmarks, while providing interpretability in the form of residue-level attention maps. In addition to being both accurate and interpretable, SPRINT is ultra-fast: querying the whole human proteome against the ENAMINE Real Database (6.7B drugs) for the 100 most likely binders per protein takes 16 minutes. SPRINT promises to enable virtual screening at an unprecedented scale, opening up new opportunities for \textit{in silico} drug repurposing and development. }

\keywords{virtual screening, structure-based drug discovery, protein language models}



\maketitle

\section{Introduction}\label{sec1}

Virtual screening has emerged as a powerful tool for predicting drug-target interactions (DTIs) and guiding experimental efforts, but conventional structure-based methods like molecular docking are often too slow for proteome-scale analyses \citep{sunseri2021virtual}. This limitation hinders their application in crucial parts of the drug discovery process such as off-target prediction \cite{singh_contrastive_2023}.
The need for scalable and interpretable virtual screening methods is particularly evident in, for example, antimicrobial drug discovery. 
The rapid emergence of antimicrobial-resistant pathogens poses a severe threat to public health \citep{world_health_organization_antimicrobial_2014}, necessitating the development of new antibiotics with novel mechanisms of action to combat cross-resistances \citep{colclough2019patterns}. 
Effective antimicrobial virtual screening requires methods that are not only fast and scalable but also interpretable, enabling researchers to: 1) identify new drug candidates with on-target effects across thousands of microbial proteomes and minimal off-target effects in humans and 2) provide interpretations for predicted DTIs and potential mechanisms of action.

Recently, vector-based virtual screening has been proposed as an alternative to structure-based screening to efficiently predict DTIs, leveraging vector featurizations for molecules \citep{morgan_generation_1965} and sequence models for protein targets \citep{brandes_proteinbert_2022, Rives2021, su_saprot_2024}.
One method, ConPLex \citep{singh_contrastive_2023}, proposes co-embedding molecules and proteins into a shared vector space, where the distance between entities is proportional to interaction likelihood. This effectively reduces the task of computing a DTI to a dot product in the co-embedding space, enabling the screening of millions of molecules against the entire human proteome in 24 hours. However, ConPLex does not scale favorably when identifying DTIs across thousands of bacterial and fungal proteomes, and it cannot provide explanations of its DTI predictions. Similarly, DrugCLIP \citep{gao2024drugclip,jia2024deep} aligns the embeddings of protein pocket structures and ligands with contrastive learning such that similarity encodes the probability of interaction.
They demonstrate state-of-the-art (SOTA) results on virtual screening benchmarks with a fraction of the compute time needed for other structure-based virtual screening methods. Their approach is restricted to structures for which binding pockets can be predicted using pocket detection algorithms or homology-based approaches; however, \cite{meller2023predicting} estimated that almost half of all structured domains may lack obvious pockets in their experimental structures.

In this work, we propose SPRINT (Structure-aware PRotein ligand INTeraction) for fast and accurate vector-based DTI predictions.
SPRINT implicitly uses structure information by featurizing proteins with SaProt \citep{su_saprot_2024}, a transformer model trained by augmenting the standard amino acid vocabulary with discrete structure-tokens \citep{van2024fast}.
Instead of averaging the per-residue embeddings from SaProt, SPRINT uses a multi-head attention pooling scheme to learn a sequence-dependent aggregation for protein representation.

SPRINT is extremely fast: querying a single protein target against the ENAMINE REAL (6.7B drugs) database and predicting its top-100 binders takes 5s when utilizing ChromaDB \citep{chroma}. Our main contributions are summarized as:
\begin{itemize}
    \item Achieving excellent performance on DTI classification (Table \ref{tab:dti_aupr}), virtual screening (Table \ref{tab:lit-pcba-results}), and binding affinity prediction (Table~\ref{tab:dti_dg}).
    \item Enhancing the second place method in the second critical assessment of computational hit-finding experiments (CACHE) challenge.
    \item Enabling pan-proteome-scale DTI screens using vector store and retrieval, scaling to billions of molecules.
    \item Improving molecular property prediction using the molecule co-embeddings learned via predicting DTIs.
\end{itemize}

Our software is available on our GitHub repository: \url{https://github.com/abhinadduri/panspecies-dti} and is also available on the web as \textit{ColabScreen}: \url{https://bits.csb.pitt.edu/colab-screen.html}.

\section{Results}\label{sec2}
\begin{table}[t]
\caption{AUPR on test sets for DTI prediction with co-embedding models across benchmarks. Results are reported as mean $\pm$ standard deviation across five independent training runs with different random seeds, evaluated on the same fixed data splits.  Train, validation, and test splits for BIOSNAP, Unseen Drugs, Unseen Targets, BindingDB, and DAVIS are taken from \citep{singh_contrastive_2023}. The MERGED dataset is split by homology (see Appendix~\ref{sec:training_details} for more details). *indicates that we did not do contrastive training on DUD-E with the ConPLex model. }
\label{tab:dti_aupr}
\bigskip
\centering

\begin{tabular}{lcccc}
\toprule
Model & ConPLex & ConPLex-attn\textsuperscript{*} & \textbf{SPRINT-sm} (10M) & \textbf{SPRINT} (16M) \\
\midrule
Backbone & ProtBert & ProtBert & SaProt & SaProt \\
Pooling & Avg & Attn & Attn & Attn \\
\midrule
BIOSNAP & 0.883$\pm$0.004 & 0.904$\pm$0.005 & \textbf{0.936}$\pm$0.001 & 0.858$\pm$0.001 \\
Unseen Drugs & 0.874$\pm$0.002 & 0.905$\pm$0.002 & \textbf{0.906}$\pm$0.001 & 0.851$\pm$0.002 \\
Unseen Targets & 0.842$\pm$0.006 & 0.844$\pm$0.011 & \textbf{0.849}$\pm$0.006 & 0.793$\pm$0.007 \\
DAVIS & 0.457$\pm$0.037 & 0.493$\pm$0.014 & \textbf{0.507}$\pm$0.005 & 0.446$\pm$0.003 \\
BindingDB & 0.616$\pm$0.009 & 0.672$\pm$0.003 & \textbf{0.718}$\pm$0.0004 & 0.588$\pm$0.0006 \\
MERGED & 0.414$\pm$0.004\textsuperscript{*} & 0.448$\pm$0.018 & 0.481$\pm$0.004 & \textbf{0.526}$\pm$0.002 \\
\bottomrule
\end{tabular}
\end{table}

\subsection{Multi-head attention pooling improves DTI prediction}
A limitation of the ConPLex framework is that it averages the per-residue embeddings obtained from PLMs \cite{singh_contrastive_2023}. As much of the relevant signal for DTIs is located in the binding pocket residues, average pooling can noise the contact map information carefully learned by PLMs through self-attention \citep{zhang2024protein}, particularly in the longer sequence-length regime. Retraining the ConPLex model with an attention-based, learned aggregation function \citep{touvron2021augmenting} achieves SOTA-competitive predictive scores for DTIs on most benchmarks (Table \ref{tab:dti_aupr}), even when using the same ProtBert model \citep{brandes_proteinbert_2022}.

To see how the learned aggregation scales with the available training data, we trained a SPRINT model on a huge dataset of DTIs from \citep{golts2024large}, which we refer to as ``MERGED'', which combines DTI data from PubChem \citep{kim2023pubchem}, BindingDB \citep{gilson2016bindingdb}, and ChEMBL \citep{gaulton2017chembl} (further details can be found in Appendix~\ref{sec:training_details}).
Our largest model, SPRINT, uses 3-layer MLPs to encode molecules and proteins after multi-head attention pooling, in contrast to SPRINT-sm's single-layer MLPs. SPRINT exhibits overfitting on the BIOSNAP, BindingDB, and DAVIS datasets but significantly improves performance on the much larger MERGED dataset (Table \ref{tab:dti_aupr}), confirming that there is value in scaling the SPRINT model size as we increase the amount of training data.

\begin{table}[tb]
\caption{Virtual Screening results on LIT-PCBA. SPRINT-ProtBert replaces SaProt with the ProtBert model, and SPRINT-Average replaces learned aggregation with average pooling and additional MLP layers. Parameter counts are shown in parentheses.}
\vspace{-2ex}
\label{tab:lit-pcba-results}
\bigskip
\centering
\begin{tabular}{ccccccc}
\toprule
 & AUROC (\%) & BEDROC (\%) & \multicolumn{3}{c}{EF} \\
\cmidrule(r){4-6}
 & & & 0.5\% & 1\% & 5\% \\
\midrule
Surflex \citep{spitzer2012surflex} & 51.47 & - & - & 2.50 & - \\
Glide-SP \cite{halgren2004glide} & 53.15 & 4.00 & 3.17 & 3.41 & 2.01 \\
Planet \citep{zhang2024planet} & 57.31 & - & 4.64 & 3.87 & 2.43 \\
\textsc{Gnina} \citep{mcnutt2021gnina} & 60.93 & 5.40 & - & 4.63 & - \\
DeepDTA \citep{ozturk2018deepdta} & 56.27 & 2.53 & - & 1.47 & - \\
BigBind \citep{brocidiacono2023bigbind} & 60.80 & - & - & 3.82 & - \\
DrugCLIP \citep{gao2024drugclip} & 57.17 & 6.23 & 8.56 & 5.51 & 2.27 \\
\midrule
SPRINT-Average (15.7M) & 67.49 & 7.80 & 7.23 & 6.26 & 3.71 \\
SPRINT-ProtBert (15.9M) & \textbf{73.4} & 11.9 & 11.68 & 10.19 & 5.27 \\
SPRINT (16M) & \textbf{73.4} & \textbf{12.3} & \textbf{15.90} & \textbf{10.78} & \textbf{5.29} \\
\bottomrule
\end{tabular}
\vspace{-3ex}
\end{table}

LIT-PCBA\citep{tran2020lit} is a challenging, commonly used virtual screening benchmark that addresses biases in the previously used DUD-E dataset \cite{wallach2018most} to explicitly enable validation of machine learning models. However, the activity labels in LIT-PCBA are derived from dose-response bioassays reported in PubChem \cite{kim2023pubchem}, which may introduce noise and variability that could potentially impact the reliability of model evaluations.
To evaluate the performance of SPRINT models at virtual screening on LIT-PCBA in the zero-shot setting, we pre-trained the deeper SPRINT (16M) model on the MERGED dataset after removing all protein sequences with $\geq$ 90\% sequence homology to the LIT-PCBA set using MMSeqs2 \citep{steinegger2017mmseqs2}.
We see that the structure-aware SPRINT models outperform competitor methods in AUROC, BEDROC (alpha = 0.85), and across all enrichment factor thresholds (Table \ref{tab:lit-pcba-results}). The SPRINT models outperform similarly sized models trained using ProtBert featurizations and multi-head attention pooling (SPRINT-ProtBert), and models trained using SaProt featurizations and average pooling (SPRINT-Average) demonstrating the importance of structure and self-attention.

We further evaluate SPRINT for the regression task of binding affinity prediction on the Therapeutic Data Commons (TDC) `BindingDB\_Patent' Leaderboard \cite{huang2022artificial}. We see in Table~\ref{tab:dti_dg} that SPRINT using ProtBert, ESM2, and SaProt PLMs is competitive with the top-ranking model on the leaderboard, Otter-Knowledge-Ensemble \cite{lam2023otter}. SPRINT-ProtBert achieves a Pearson correlation of 0.593, compared with 0.588 for Otter-Knowledge-Ensemble and 0.598 in our evaluation of DTIAM \cite{lu2025dtiam}. This is notable, as Otter-Knowledge-Ensemble is an ensemble of four knowledge graph-refined protein and ligand representations, with each set of representations fine-tuned on a separate dataset of protein-ligand interactions, while DTIAM combines pretrained molecular and protein representations with an AutoML ensemble. The SPRINT models achieve comparable performance while only trained on the BindingDB\_Patent training set. ProtBert achieves the highest mean correlation among the SPRINT variants; SaProt and ESM2 achieve comparable performance.

\begin{table}[t]
\caption{Pearson’s correlation coefficient on the TDC BindingDB\_Patent Leaderboard. Results are reported as mean $\pm$ standard deviation across five independent training runs with different random seeds, evaluated on the same data splits.}
\label{tab:dti_dg}
\bigskip
\centering
\begin{tabular}{cccccccc}
\toprule
Model & Pearson's Correlation Coefficient \\
\midrule
Otter-Knowledge-Ensemble\cite{lam2023otter} & $0.588\pm0.002$ \\
DTIAM \cite{lu2025dtiam} & $0.598\pm0.0044$\\
\midrule
SPRINT-ProtBert & $0.593\pm0.005$ \\
SPRINT-ESM2 & $0.582\pm0.012$ \\
SPRINT & $0.588\pm0.011$ \\
\bottomrule
\end{tabular}
\end{table}

\subsection{Enhancing Real-World Virtual Screening in CACHE2}
\begin{figure}
    \centering
    \includegraphics[width=0.5\linewidth]{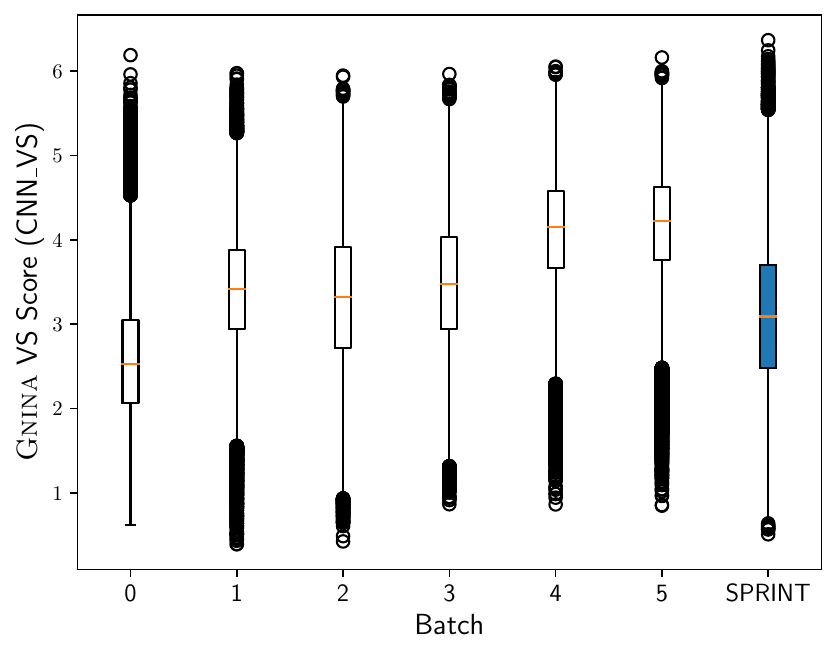}
    \caption{\textsc{Gnina} CNN\_VS scores \cite{sunseri2021virtual} (the product of the GNINA predicted affinity and pose score) of the molecules found during DeepDocking and the molecules picked with SPRINT. Higher CNN\_VS scores indicate a greater likelihood of interaction.}
    \label{fig:deepdocking-batch}
\end{figure}

The Critical Assessment of Computational Hit-Finding Experiments (CACHE) \cite{ackloo2022cache} serves as a benchmark for prospective evaluation of virtual screening methods through experimental validation of predicted hits. CACHE challenge 2 focused on the RNA binding site of SARS-CoV-2 NSP13, representing one of the most conserved pockets in SARS-CoV-2 and across coronaviruses \cite{yazdani2021genetic} with no known inhibitors. Other efforts have predicted inhibitors for the ATP binding site of the NSP13 helicase\cite{el2022multi} which would similarly result in reduced RNA replication, however, this pocket is significantly less conserved across coronaviruses. Crystal structures with bound fragments in the RNA-binding site exist, with PDB 5RLZ \cite{newman2021structure} used for virtual screening. The second-place team \cite{cache2_method} implemented a DeepDocking \cite{gentile2020deep} approach using \textsc{Gnina} \cite{sunseri2021virtual,mcnutt2021gnina}. They first filtered the Enamine REAL database for drug-like properties: molecules were removed if they were considered a pan-assay interference compound, logp was greater than 4, molecular weight was greater than 500 Da, or heavy atom count was less than 14. Filtering reduced the dataset to 4.4 billion molecules. A random sample of 100,000 molecules (Batch 0) was docked against the 5RLZ structure using \textsc{Gnina}, and a surrogate model was trained to predict the docking score directly from a fingerprint of the molecule \cite{morgan_generation_1965}. The entire database was scored with the surrogate model, and the top-scoring 100,000 molecules (Batch 1) were docked again with \textsc{Gnina} to produce training data for a new surrogate model. This process was repeated until the docking scores converged (Batch 5), as shown in Figure~\ref{fig:deepdocking-batch}.

To evaluate the utility of SPRINT in filtering a compound database for high-likelihood binders and accelerating the DeepDocking process, we used SPRINT, pretrained on the MERGED dataset, to predict DTI scores for the filtered Enamine REAL database against the same virtual screening structure (5RLZ). We selected the highest-scoring 100,000 molecules from the filtered database and docked them using \textsc{Gnina}, following the same docking protocol as \cite{cache2_method}.  The virtual screening scores (CNN\_VS \cite{sunseri2021virtual}, the product of the GNINA predicted affinity and pose score) of our docked compounds are compared to the batches of \textsc{Gnina} DeepDocking in Figure~\ref{fig:deepdocking-batch}. Molecules selected by SPRINT had a higher average CNN VS score (3.10) than Batch 0 of DeepDocking (2.59), and are more distributionally similar to Batches 1, 2, and 3. Notably, SPRINT finds 16 diverse, high-scoring molecules (CNN VS $>$ 6) shown in Figure~\ref{fig:SPRINT_molecules}, while only 6 high-scoring molecules are found through DeepDocking (with one high-scoring molecule found during the initial random selection, Batch 0). 

\subsection{SPRINT enables querying binding partners from 5132 proteomes}
\begin{figure}[tb]
    \centering
    \includegraphics[width=8cm]{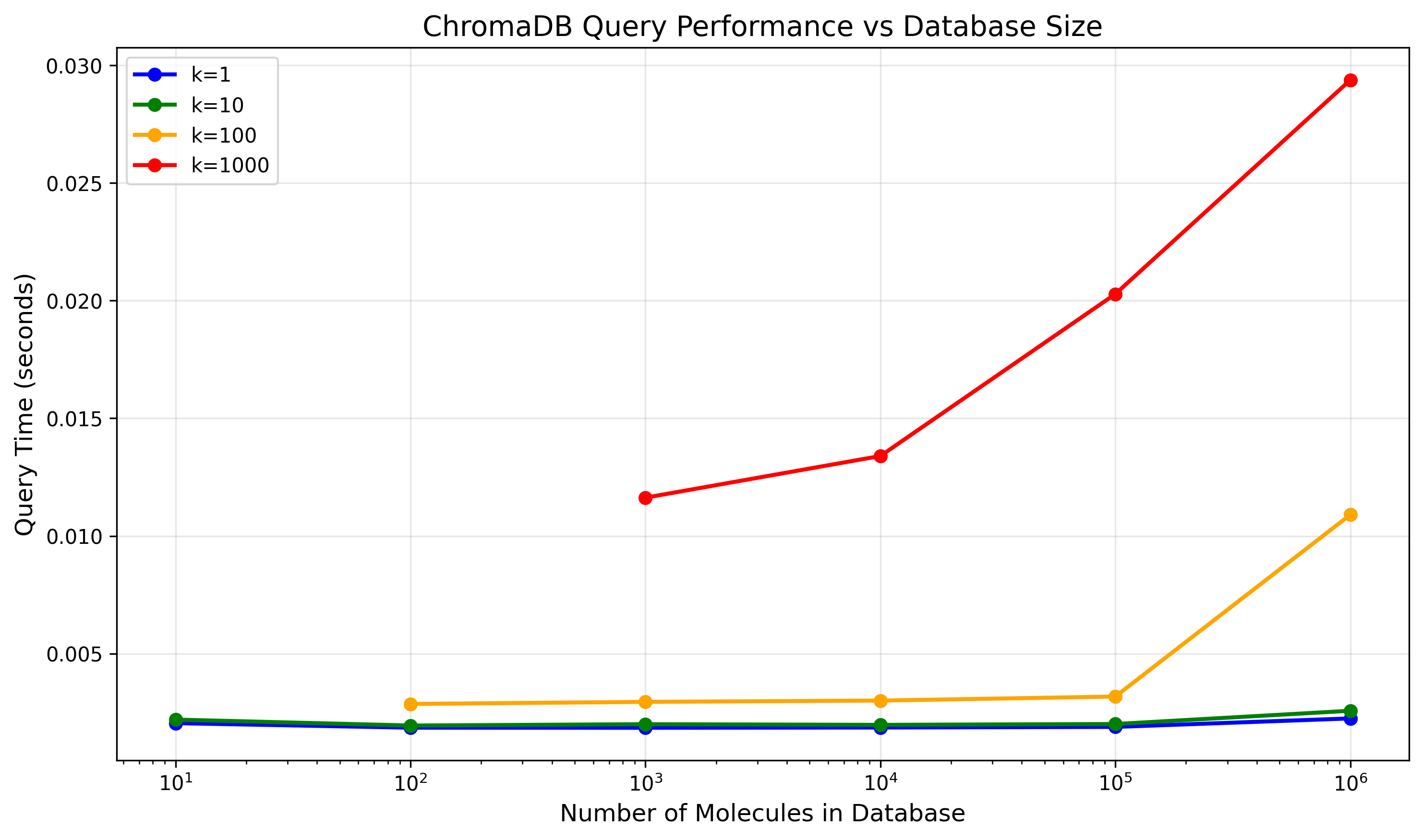}
    \caption{Times for predicting the top DTIs for a ligand using vector search.}
    \label{fig:query_times}
\end{figure}

To demonstrate the utility of SPRINT at the pan-species-proteome scale, we constructed a dataset of 5,043 bacterial proteomes, 88 fungal proteomes, and the human proteome, containing 4,291,525 total protein sequences.
To store and query the co-embeddings, we use the Chroma vector store\cite{chroma},
a tool developed for semantic search and retrieval-augmented generation in natural language processing. The scaling properties of this framework are highly favorable (Fig. \ref{fig:query_times}): 
querying a ligand for the 100 most likely binders against the entirety of UniProt (60M sequences) takes about 0.5s, and querying all $2\mathrm{e}6$ molecules in CHEMBL for each of their 10 most likely binders against the 4.3M proteins in our multi-species dataset takes less than 3 hours. All timings were calculated on an otherwise unloaded 48-core 2.5 GHz Intel Xeon E5-2680 machine.

\begin{table}[t]
\caption{
Performance of MLP models using different molecular featurizations. Results are reported as mean $\pm$ standard deviation across five independent training runs with different random seeds, evaluated on the same  data splits. Higher is better for F1 and AUC; lower is better for MSE. Models trained on Morgan fingerprints used a larger hidden size to match the size of models trained on fingerprints concatenated with embeddings. Unless otherwise specified, benchmarks are taken from MoleculeNet. \cite{wu2018moleculenet}.}
\label{tab:propertypred}
\bigskip
\centering
\small

\begin{tabular*}{\linewidth}{@{\extracolsep{\fill}}lcc@{}}
\toprule
\multicolumn{3}{c}{\textbf{Classification (F1 $\uparrow$)}} \\
\cmidrule(lr){1-3}
Featurization & Antibacterial \cite{terlouw2023mibig} & Toxicity \cite{setiya2024moltoxpred} \\
\midrule
Morgan Fingerprint
    & 0.740 $\pm$ 0.027
    & 0.720 $\pm$ 0.008 \\

SPRINT-sm Embedding
    & 0.687 $\pm$ 0.012
    & 0.656 $\pm$ 0.015 \\

Morgan + SPRINT-sm 
    & \textbf{0.749} $\pm$ 0.016
    & \textbf{0.735} $\pm$ 0.006 \\

SPRINT Embedding
    & 0.614 $\pm$ 0.027
    & 0.631 $\pm$ 0.023 \\

Morgan + SPRINT 
    & 0.722 $\pm$ 0.027
    & 0.701 $\pm$ 0.018 \\
\bottomrule
\end{tabular*}

\vspace{1.5em}

\begin{tabular*}{\linewidth}{@{\extracolsep{\fill}}lccc@{}}
\toprule
\multicolumn{4}{c}{\textbf{Classification (AUC $\uparrow$)}} \\
\cmidrule(lr){1-4}
Featurization & Tox21 & SIDER & ToxCast \\
\midrule
Morgan Fingerprint
    & 0.606 $\pm$ 0.025
    & 0.608 $\pm$ 0.035
    & 0.471 $\pm$ 0.053 \\[1mm]

SPRINT Embedding
    & 0.681 $\pm$ 0.028
    & 0.555 $\pm$ 0.038
    & 0.553 $\pm$ 0.055 \\[1mm]

\makecell[l]{Morgan + SPRINT }
    & \textbf{0.751} $\pm$ 0.026
    & \textbf{0.620} $\pm$ 0.038
    & \textbf{0.606} $\pm$ 0.055 \\
\bottomrule
\end{tabular*}

\vspace{1.5em}

\begin{tabular*}{\linewidth}{@{\extracolsep{\fill}}lccc@{}}
\toprule
\multicolumn{4}{c}{\textbf{Regression (MSE $\downarrow$)}} \\
\cmidrule(lr){1-4}
Featurization & Delaney & FreeSolv & Lipophilicity \\
\midrule
Morgan Fingerprint
    & 0.646 $\pm$ 0.029
    & 0.403 $\pm$ 0.113
    & 0.640 $\pm$ 0.025 \\[1mm]

SPRINT Embedding
    & 0.862 $\pm$ 0.046
    & 0.415 $\pm$ 0.130
    & 0.864 $\pm$ 0.043 \\[1mm]

\makecell[l]{Morgan  + SPRINT }
    & \textbf{0.612} $\pm$ 0.026
    & \textbf{0.378} $\pm$ 0.127
    & \textbf{0.611} $\pm$ 0.022 \\
\bottomrule
\end{tabular*}

\end{table}

\subsection{Pre-training to predict DTIs improves property prediction}
To benchmark the usefulness of DTI co-embeddings for marginal property prediction, e.g., predicting the properties of a compound only from its molecular graph, we computed SPRINT DTI co-embeddings for several benchmarks of molecular property prediction\citep{setiya2024moltoxpred,terlouw2023mibig,wu2018moleculenet}. 
As shown in Table \ref{tab:propertypred}, concatenating SPRINT-sm embeddings with Morgan fingerprints improved mean F1 on both antibacterial and toxicity classification tasks. Concatenating SPRINT embeddings with fingerprints improved mean performance on all six additional MoleculeNet tasks, although it reduced mean F1 on the antibacterial and toxicity tasks. Embeddings alone generally underperformed fingerprints, with exceptions on Tox21 and ToxCast.  These results suggest that DTI-derived molecular embeddings can complement traditional fingerprints, although their utility for downstream property prediction depends on the task and the pre-training model. The stronger performance of SPRINT-sm on antibacterial and toxicity classification suggests that increasing model capacity for DTI prediction does not necessarily improve the transferability of the learned molecular representation.

\section{Methods}\label{sec11}

To enable fast and accurate screens, we seek a co-embedded representation of drugs and protein targets where a simple similarity metric indicates binding likelihood.
Let $D$ and $T$ denote the random variables representing drugs and targets, $f$ and $g$ denote the choice of frozen drug and target encoders, and $C_d$ and $C_t$ denote modality-specific neural networks that project drug and target embeddings, respectively, into a shared co-embedding space. Let $Y$ denote the random variable representing drug-target interaction, where $Y = 1$ denotes an interacting pair, and $Y = 0$ denotes a non-interacting pair. Denoting latent co-embeddings $Z_d = C_d(f(D))$ and $Z_t = C_t(g(T))$, our model is:

\begin{equation}
P(Y = 1 \mid Z_d, Z_t)
= \sigma\left(
\alpha 
\underbrace{\frac{Z_d \cdot Z_t}
{\lVert Z_d\rVert \lVert Z_t\rVert}}_
{\operatorname{\mathrm{cos\_sim}}(Z_d,Z_t)}
\right),
\label{main_eq}
\end{equation}


where $\sigma$ denotes the sigmoid activation function, and $\alpha$ is a constant scaling factor chosen to saturate the range of the sigmoid function, as unscaled cosine similarity ranges from $(-1, 1)$. In our implementation, we choose $\alpha=5$. Our goal through training is to learn $C_d$ and $C_t$ that minimize binary cross-entropy loss against ground truth binding and non-binding pairs.

In addition to classifying drug-target binding, SPRINT is capable of predicting binding affinity without changing the architecture of the model. By replacing the cosine similarity with a dot-product and removing the final sigmoid, we can leverage SPRINT as a binding affinity predictor (Equation~\ref{eq:affinity}):
\begin{equation}
    p\hat{K} = \left( {Z_d} \cdot {Z_t} \right)
    \label{eq:affinity}
\end{equation}

where $p\hat{K}$ is the predicted $pK=-log_{10}(K_d)$ to simplify training of the model on affinity values. Binding affinity regression models are trained to minimize mean square error loss against ground truth binding affinity by updating the weights of $C_d$ and $C_t$.

\begin{figure}
    \centering
    \includegraphics[width=.9\textwidth]{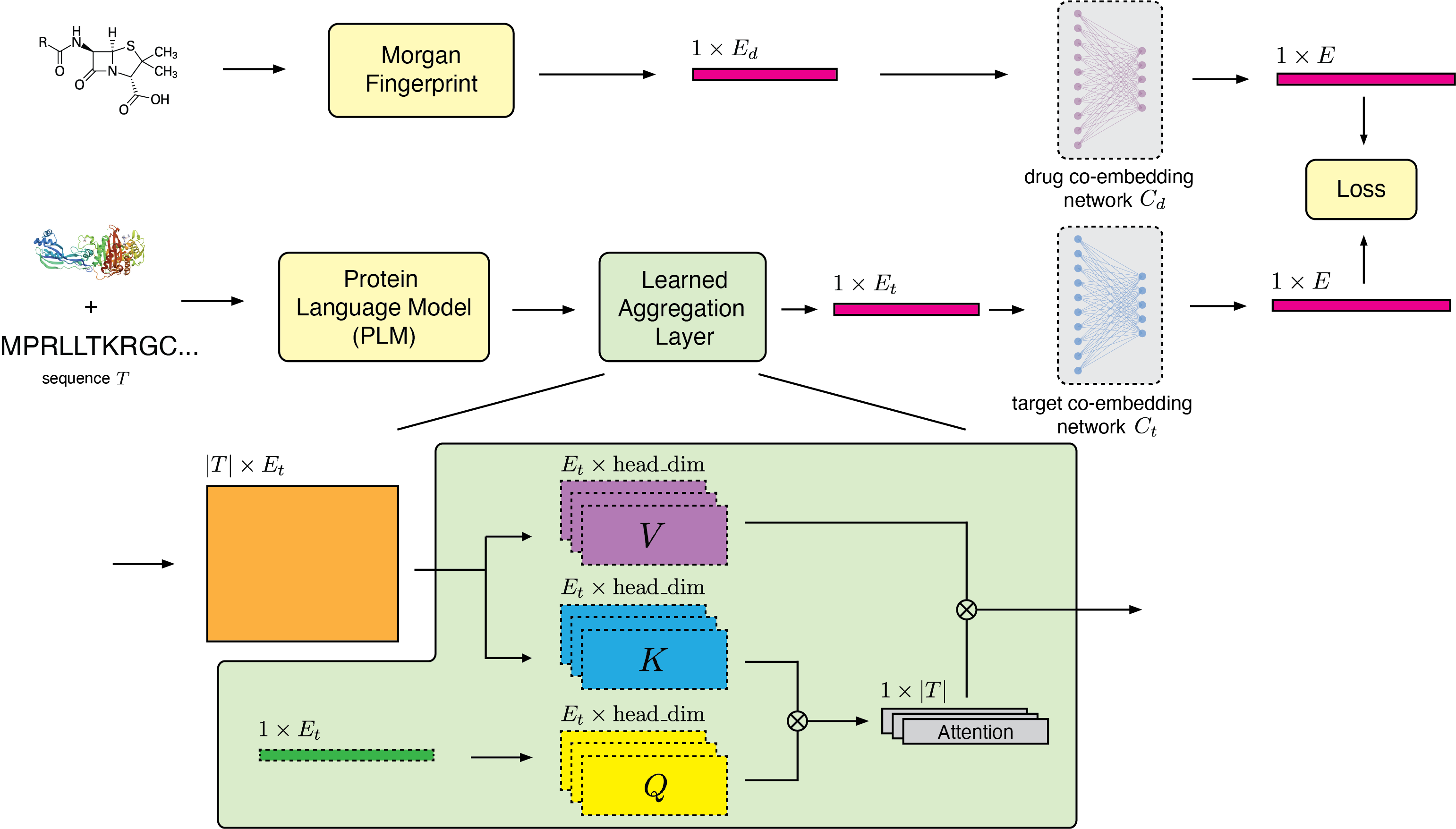}
    \caption{SPRINT learns protein representations via a multi-head attention pooling scheme. Then, SPRINT learns a shared co-embedding space between molecules and protein targets via modality-specific neural networks $C_d$ and $C_t$. The model is trained end-to-end via a binary cross entropy loss on binding and non-binding drug-target pairs, where the probability of interaction is computed as a sigmoid function of the cosine distance between the drug and target embeddings. The learnable parameters of the network are depicted with dashed borders.}
    \label{fig:sprint-fwd}
\end{figure}

For the drug encoder $f$, we use the Morgan fingerprint featurizer available in RDKit with bit length 2048 and radius 2 \citep{rdkit, morgan_generation_1965}. For the target encoder $g$, we choose the structure-aware transformer model SaProt \citep{su_saprot_2024}, a structure-aware protein language model that outputs per-residue embeddings, resulting in a $|T| \times E_t$ featurization for an input sequence $T$. SaProt optionally takes protein structure as an input to compute FoldSeek tokens for embedding \citep{van2024fast}. We utilize AlphaFold2 \citep{jumper2021highly} predicted structures to generate structure tokens when training the DTI model with SaProt. Unlike prior works, we employ multi-head attention pooling to aggregate these per-residue embeddings into a single vector representation of a protein (Fig. \ref{fig:sprint-fwd}). This approach has two merits. First, we hypothesize that the model will be able to focus on information-rich residues due to the data-dependent nature of the attention scheme. Second, we can gain insights into the biological relevance of the attention patterns learned by the model by analyzing which residues are prioritized and how they may relate to known mechanisms of drug-target interaction, an avenue which is explored in Appendix~\ref{sec:Attn_interrogation}. Further training details and hyperparameters are provided in Appendix~\ref{sec:training_details}.

\subsection{Data}
We evaluate the performance of the SPRINT DTI prediction architecture on two tasks: interaction prediction and binding affinity prediction. SPRINT models trained for interaction prediction predict 1 if a drug and target pair interact and 0 otherwise. We utilize the same interaction prediction datasets as \cite{singh_contrastive_2023}: DAVIS, BIOSNAP, and BindingDB. DAVIS\cite{davis2011comprehensive} and BindingDB\cite{liu2007bindingdb} are both composed of drug-target paired with experimentally annotated dissociation constants, $K_d$. We utilize the same thresholds as \citep{singh_contrastive_2023} to set the binary labels: $K_d<30$ are labeled as interacting and $K_d\geq30$ are non-interacting. DAVIS consists of only 2086 interactions, BindingDB has 12668, and BIOSNAP has 19238 DTIs. The BIOSNAP\cite{zitnik2018biosnap} data originally consisted of only interacting drug-target pairs, so negative pairs are created by randomly sampling an equal number of drug-target pairs. The Unseen Drugs and Unseen Targets datasets are different splits of BIOSNAP from \cite{huang2021moltrans} to elucidate the zero-shot performance of DTI methods. The Unseen Targets dataset is constructed by randomly selecting 20\% of the proteins from the full BIOSNAP dataset and assigning all interactions involving these proteins to the test set, while the remaining interactions are used for training. The Unseen Drugs dataset is constructed in an identical manner, but using the drugs rather than the targets.

All dataset splits are the same as \citep{huang2021moltrans}, 70/10/20 for training, validation, and test, respectively. The training split is artificially subsampled to have an equal number of positive and negative interactions. To evaluate the scaling of the model on more training tokens, we utilize the ``MERGED'' dataset constructed by \cite{golts2024large} as a DTI dataset for benchmarking large model performance. This dataset is composed of data from PubChem \citep{kim2023pubchem} (98.31\%), BindingDB \citep{gilson2016bindingdb} (1.17\%), and ChEMBL \citep{gaulton2017chembl} (0.52\%). We use the activity label of the dataset for our interaction label providing 929,656 positive pairs and 83,703,190 negative pairs. We split the data into train, validation, and test splits such that targets in train and validation have at most 90\% sequence homology, and targets in train and test have at most 70\% sequence homology (splits are further detailed in Appendix~\ref{sec:training_details}).

We use the Therapeutic Data Commons (TDC) \cite{huang2022artificial} BindingDB\_Patent dataset for binding affinity prediction. Each drug-target interaction is labeled with an experimentally determined $IC_{50}$. The BindingDB\_Patent dataset splits train and test with a temporal split on the patent date, everything before 2019 is included in the train and validation split, while everything 2019 and after is in the test set. Temporal splits are still prone to data leakage as new drugs are commonly developed for established protein targets as well as drugs being reused for new targets \cite{li2024leak}. The TDC hosts a leaderboard for this dataset to continuously evaluate new methods on the same training and test splits. We reserve a random 20\% of the training set as a validation set.


\section{Discussion}

Vector-based screening enables rapid DTI prediction at scales that are computationally impractical for conventional structure-based approaches. SPRINT advances this framework through multi-head attention pooling, with larger models benefiting from increased DTI training data. SPRINT achieves performance comparable to state-of-the-art methods for DTI classification and virtual screening, and matches Otter-Knowledge-Ensemble\citep{lam2023otter}, the top method on the TDC BindingDB\_Patent leaderboard, for binding affinity prediction without explicit knowledge graphs of protein–ligand interactions. Its co-embedding strategy supports rapid screening through sublinear-time retrieval of predicted binding partners, enabling searches across proteomes from thousands of species, including applications to antimicrobial discovery (Figure~\ref{fig:coembedding_umap}). Finally, learning molecular co-embeddings through DTI prediction provides an effective pre-training strategy that improves molecular property prediction when these embeddings are combined with traditional fingerprints (Table~\ref{tab:propertypred}).

We also show that structure-aware PLMs like SaProt can confer performance gains in virtual screening.  We compare the virtual screening performance of SPRINT on the second CACHE challenge to the second place computational method which utilized DeepDocking.  SPRINT enables direct querying of the entire ENAMINE database. After performing only 1/6th of the docking, SPRINT finds almost three times the number of high-scoring molecular scaffolds compared to {\textsc{gnina}} DeepDocking. SPRINT's top molecules can be a useful starting point for DeepDocking to ensure that chemical space is more effectively and efficiently explored. 

We envision the SPRINT framework and training task as a useful benchmarking tool for protein and molecule encoders. Future work will evaluate other structure-aware PLMs such as MULAN or S-PLM \cite{frolova2024mulan,Wang2023.08.06.552203}, and other pre-trained molecule encoders like UniMol \cite{zhou2023uni,ji2024uni}, although initial experimentation does not find such embeddings to be superior to traditional fingerprints (Appendex~\ref{sec:training_details}). Additional work to interpret the aggregated protein features through methods like InterPLM\citep{simon2024interplm} could provide insight into the virtual screening predictions. We anticipate this technology will democratize virtual screening by enabling the discovery of lead molecules with a fraction of the compute cost required by comparable, structure-based, virtual screening methods.

\backmatter

\bmhead{Supplementary information}

Additional information provided in the appendices.

\bmhead{Acknowledgments}

The authors would like to thank Ian Dunn for providing the filtered dataset used for the CACHE screen in addition to the results of their virtual screening method for comparison.


\section*{Declarations}

\subsection*{Funding}
A.M. and D.K. are supported by the R35GM140753 from the National Institutes of Health. 
A.A. and H.M. are supported by the United States Department of Agriculture, Agricultural Research
Service (USDA-ARS) and by grants 5R01GM107550-10 and 1U24DK133658-01.
C.E. and E.X. are supported by the R01GM140467 from the National Institutes of Health, and the 2040381 from the National Science Foundation.
M.D. was supported by the F31LM014194 from the National Institutes of Health.

\subsection*{Competing interests}
The authors declare no competing interests.

\subsection*{Data availability}
DAVIS, BIOSNAP, BindingDB, and MERGED DTI classification datasets are available through our github: \url{https://github.com/abhinadduri/panspecies-dti}. The BindingDB DTI DG dataset for the TDC Leaderboard is available through the Therapeutics Data Commons: \url{https://tdcommons.ai/}. The Lit-PCBA evaluation benchmark is available through the original authors \url{https://drugdesign.unistra.fr/LIT-PCBA/}. The Enamine REAL dataset is available from \url{https://enamine.net/}.

\subsection*{Code availability}
All code to reproduce our results is available on our github repository: \url{https://github.com/abhinadduri/panspecies-dti}.

\subsection*{Author contribution}
\textbf{A.M.}: conceptualization, methodology, writing-original draft, software, investigation, validation
\textbf{A.A.}: conceptualization, methodology, writing-original draft, software, investigation, validation
\textbf{C.E.}: conceptualization,  methodology, software, writing-original draft
\textbf{N.D.} software, visualization, writing-review\&editing
\textbf{T.K.} software, visualization, writing-review\&editing
\textbf{M.D.}: software, visualization, writing-review\&editing
\textbf{E.P.}: supervision, funding acquisition
\textbf{H.M.}: supervision, funding acquisition
\textbf{D.K.}: conceptualization, resources, supervision, funding acquisition, writing-review\&editing

\newpage
\begin{appendices}
\appendix
\counterwithin*{table}{section}
\counterwithin*{figure}{section}

\section{SPRINT recovers known mechanisms of action}
\begin{figure}
    \centering
    \includegraphics[width=\textwidth]{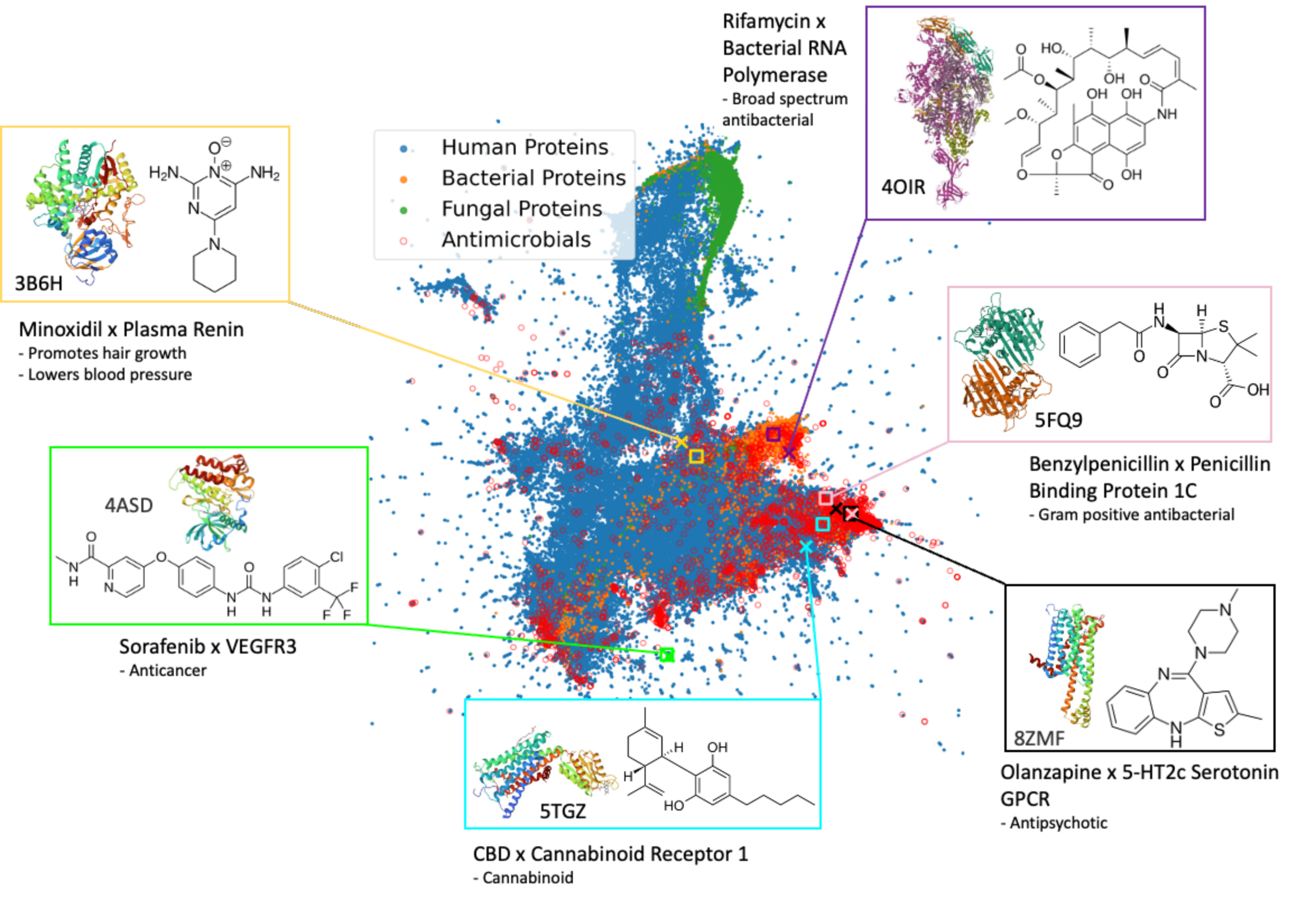}
    \caption{UMAP visualization of the binding co-embedding space of drug-like small molecules and their protein targets across bacterial, fungal, and human proteomes. We see that antimicrobial compounds co-localize with regions of the shared latent space that contain human, bacterial, and fungal proteomes.}
    \label{fig:coembedding_umap}
\end{figure}

Our pan-species protein dataset is comprised of all predicted protein sequences in reference genomes from NCBI within the taxons bacteria, fungus, and human.
Each taxon contained 3,379,854, 775,477, and 136,194 protein sequences respectively. We gathered a list of 3,112 natural products \cite{terlouw2023mibig}, as they are known to have a high prior likelihood for antimicrobial activity \citep{atanasov2021natural} and are out-of-distribution relative to our MERGED training dataset. We then co-visualized several antimicrobials and drugs with their known protein targets across microbial proteomes (Fig. \ref{fig:coembedding_umap}), recovering several known mechanisms of action. The dataset is available to query at \url{https://bits.csb.pitt.edu/colab-screen.html}.

\section{Training details}
\label{sec:training_details}
DTI models are trained using the same train/val/test splits as \cite{singh_contrastive_2023} for the DAVIS, BindingDB, and BIOSNAP datasets. All structure tokens for SaProt were computed on AlphaFold2 \citep{jumper2021highly} generated structures. Structures were downloaded from the AlphaFold Protein Structure Database if they existed. When no precomputed structure was available, ColabFold \citep{mirdita2022colabfold} was run with 2 random seeds to generate 10 energy minimized structures, and the minimized structure with the highest pLDDT was used. Structure tokens were generated with Foldseek \citep{van2024fast}, masking the structure token if the residue pLDDT was less than 70.

Proteins are encoded using a frozen, pretrained PLM, $g$, which outputs a matrix, $g(T)\in \mathds{R}^{|T|\times E_t}$ encoding each residue into a vector of length $E_t$ for the length of the sequence $|T|$. Rather than using an average across the sequence like prior works \cite{singh_contrastive_2023}, we employ a multi-head attention pooling layer that we call the ``Learned Aggregation Layer'' which is constructed of a slightly modified attention mechanism. Similar to standard attention, the keys, $K$, and values, $V$, are linear maps from the PLM embedding of the protein while the query,$Q$ is computed from a learned, input-independent quantity, $\mathrm{c}\in\mathds{R}^{1\times E_t}$. The queries, keys, and values are derived through equation \ref{eq:qkv}.
\begin{align}
\label{eq:qkv}
    Q &= \mathrm{c}W_Q\quad&\in\mathds{R}^{H\times1\times d}\notag\\
    K &= g(T)W_K\quad&\in\mathds{R}^{|T|\times H\times d}\\
    V &= g(T)W_V\quad&\in\mathds{R}^{|T|\times H\times d}\notag
\end{align}
Where $W_Q,W_K,W_V$ are all learned weight matrices in $\mathds{R}^{E_t\times E_t}$ and followed by a reshape into $H$ attention heads of dimension $d=E_t/H$. This is mathematically equivalent  to using independent, per-head projection matrices $W_Q^{(h)},W_K^{(h)},W_V^{(h)}\in\mathds{R}^{d\times E_t}$ where for each head we would compute:
\begin{align*}
Q^{(h)}=cW_Q^{(h)}\\
K^{(h)}=g(T)W_K^{(h)} \\
V^{(h)}=g(T)W_V^{(h)}
\end{align*}
Attention, $A$, is computed as the softmax of the product between the queries and keys.
\begin{equation}\label{eq:attn}
    A=\mathrm{softmax}(QK^T)\quad\in\mathds{R}^{H\times1\times|T|}
\end{equation}
The output vector, $t$, representing the entire length of the protein is the product of the attention matrix, $A$, and the values, $V$, followed by a final linear mapping.
\begin{equation}
    t = (AV)W_O+ b_O \quad\in\mathds{R}^{1\times E_t}
\end{equation}
Where $W_O\in\mathds{R}^{E_t\times E_t}$ and $b_O\in\mathds{R}^{1\times E_t}$. This learned aggregation layer is followed by a feed forward network to map the protein embedding, $t$, into the co-embedding space $Z$. The combination of the learned aggregation layer and the feed forward network are termed $C_t$ and are learned during training.

During training, the loss $\mathcal{L}$ is written as

\begin{equation}
    \mathcal{L} = \frac{1}{N}\sum\limits_i^N \left[ Y^i\log(\tilde{Y}^i) + (1-Y^i)\log(1-\tilde{Y}^i) \right]
    \label{eq:loss}
\end{equation}
\begin{equation*}
    \tilde{Y}^i = P(Y^i = 1 | Z_d^i, Z_t^i) = \sigma\left(\alpha \, \frac{Z_t^i}{\left\| Z_t^i\right\|}\cdot\frac{Z_d^i}{\left\|Z_d^i\right\|}\right)
    \label{eq:sim}
\end{equation*}

where the protein, $T^i$, and drug, $D^i$, have been mapped to the SPRINT co-embedding space as $Z_t^i$ and $Z_d^i$, respectively. $Y_i\in\{0,1\}$ is a ground-truth label with value 1 if $T^i$ and $D^i$ are binders or 0 if they are non-binders. The pre-sigmoid scalar value, $\alpha$, is used to expand the range of cosine-similarity to the domain of the sigmoid. We set $\alpha$ to 5. We explored several co-embedding dimensions for $Z$ in Table \ref{tab:dim_coemb} and find that the dimension of the co-embedding space does not substantially affect model performance, we therefore train all models with $Z=1024$ for closer comparison with \cite{singh_contrastive_2023}. Models are trained for binary classification using standard, supervised cross-entropy loss in equation \ref{eq:loss}.

Binding affinity regression models use the same architecture to learn the co-embedding space, but the final, non-learned components are removed. The sigmoid and pre-sigmoid scalar value are removed and the cosine similarity is replaced with a dot-product (Equation~\ref{eq:affinity}). This allows the magnitude of drug and target embeddings to impact the prediction as well as increasing the range to all real numbers. Regression models are trained with mean squared error loss.

\begin{table}[b]
    \caption{Dimension of co-embedding space and affect on DTI classification performance}
    \label{tab:dim_coemb}
    \centering
    \begin{tabular}{cccc}
    \toprule
        Dimension & BindingDB & BIOSNAP & DAVIS \\
    \midrule
         512 & 0.721 $\pm$ 0.001 & 0.937 $\pm$ 0.001 & 0.521 $\pm$ 0.008 \\
         \textbf{1024} & 0.722 $\pm$ 0.004 & 0.939 $\pm$ 0.000 & \textbf{0.523} $\pm$ 0.011 \\ 
         2048 & \textbf{0.727} $\pm$ 0.001 & \textbf{0.941} $\pm$ 0.001 & 0.520 $\pm$ 0.010 \\
    \bottomrule
    \end{tabular}
\end{table}

\begin{table}[b]
    \caption{DTI Performance of SPRINT compared to DTIAM. On these small datasets, SPRINT is prone to overfitting, but competitive performance to DTIAM is observed with the smaller model.}
    \label{tab:dtiam}
    \centering
    \begin{tabular}{cccc}
    \toprule
        Model & BindingDB & BIOSNAP & DAVIS \\
    \midrule
         DTIAM & 0.723 $\pm$ 0.0057 & 0.952 $\pm$ 0.0015 & 0.530  $\pm$ 0.0125 \\
         {SPRINT-sm (10M)} & 0.718 $\pm$ 0.0004 & 0.936 $\pm$ 0.001 & 0.507 $\pm$ 0.005 \\ 
         {SPRINT} (16M) & 0.588 $\pm$ 0.0006 & 0.858 $\pm$ 0.001 & 0.446 $\pm$ 0.003 \\
    \bottomrule
    \end{tabular}
\end{table}

ConPLex models are trained with the hyperparameters used in the original paper \cite{singh_contrastive_2023}. Our Attention Pooling models are trained with a learning rate of $1\times10^{-5}$ and a dropout value of 0.05 for 250 epochs, keeping all other hyperparameters the same as ConPLex training. We use four heads in our learned aggregation layer as this showed competitive performance across the BindingDB, BIOSNAP, and DAVIS datasets (Table \ref{tab:agg_heads}) We set weight decay to 0.01 for the binding affinity regression model following hyperparameter tuning on the validation set. The model checkpoint with the highest validation AUPR for classification models and lowest MSE for regression models during training is evaluated on the test set (Table \ref{tab:dti_aupr} and~\ref{tab:dti_dg}).  Unless otherwise stated, results reported as mean $\pm$ standard deviation summarize five independently trained downstream models with different random seeds, evaluated on the same fixed data splits.

\begin{table}[t]
    \caption{Changing the number of heads in the learned aggregation layer and affect on DTI classification performance (AUPR, mean$\pm$std)}
    \label{tab:agg_heads}
    \centering
    \begin{tabular}{cccc}
    \toprule
         Heads in Agg Layer & BindingDB & BIOSNAP & DAVIS \\
    \midrule
         1 & 0.7218 $\pm$ 0.002 & \textbf{0.9408} $\pm$ 0.001 & 0.515 $\pm$ 0.004 \\
         \textbf{4} & \textbf{0.7223} $\pm$ 0.003 & 0.9389 $\pm$ 0.001 & 0.521 $\pm$ 0.009 \\
         8 & 0.7217 $\pm$ 0.003 & 0.9388 $\pm$ 0.001 & \textbf{0.525} $\pm$ 0.012 \\
    \bottomrule
    \end{tabular}
    
\end{table}

To enable efficient training on the MERGED dataset, we featurize the unique proteins and molecules before training, storing their representations in memory-mapped files for quick retrieval using the Lightning Memory-Mapped Database Manager (LMDB) library. To address data imbalance in the binding data, for each epoch, we train using all of the drug-target binding pairs, and subsample an equivalent number of non-binding pairs without replacement. We observed that models trained with more negatives than positives (at a 3:1 ratio), achieved better virtual screening performance, but had less interpretable attention patterns (Tables \ref{tab:dti_aupr}, \ref{tab:noneg_sample}). Models are trained for 20 epochs. All other hyperparameters were kept the same.

Our method leverages an asymmetric architecture for protein and small molecule encoding due to our use of PLM encoding and Morgan Fingerprint, respectively. The Morgan Fingerprint is an easy to calculate, hashed fingerprint widely used in cheminformatics that shows remarkably strong performance across a range of tasks, often outperforming more complex neural network approaches \cite{praski2025benchmarking}. We investigated two other molecular fingerprints within the SPRINT framework: MoLFormer \cite{ross_large-scale_2022} and CLAMP\cite{seidl2023enhancing}(Table \ref{tab:Molablate}). MoLFormer is a self-supervised SMILES transformer model that is pretrained using a masked language modeling objective, similar to the PLMs. We used the provided pre-trained checkpoints which were trained on 10\% of ZINC and 10\% of Pubchem, about 100 million molecules utilizing the average of the molecule encoding as $f(D)$. CLAMP uses an alternative supervised contrastive learning scheme, similar to SPRINT, in which molecules are paired with ChEMBL bioactivity assay data, incorporating a bioactivity-specific inductive bias into the fingerprint derived representation. The pretrained CLAMP model was frozen and utilized for generating the molecule encoding, $f(D)$. We find that MolFormer encoding performs about the same as Morgan Fingerprint of the molecules across the DAVIS, BindingDB, and BIOSNAP datasets (Table \ref{tab:Molablate}) at a much higher computational cost. CLAMP molecule encodings decrease performance on the same datasets.

\begin{table}[bh]
    \caption{Ablating molecule encoder and effect on DTI classification performance, SPRINT-sm}
    \label{tab:Molablate}
    \centering
    \begin{tabular}{cccc}
    \toprule
         Molecule Encoder & DAVIS & BIOSNAP & BindingDB \\
    \midrule
        \textbf{Morgan Fingerprint} & 0.524$\pm$0.009 & 0.939$\pm$0.0004 & 0.722$\pm$0.003 \\
        MoLFormer & 0.524$\pm$0.009 & 0.934$\pm$0.0004 & 0.715$\pm$0.003 \\
        CLAMP & 0.496$\pm$0.009 & 0.932$\pm$0.0007 & 0.707$\pm$0.004 \\
    \bottomrule    
    \end{tabular}
    
\end{table}

Finally, we explore the importance of the PLM within the SPRINT framework. PLMs are trained on millions of sequences utilizing a masked language modeling approach\cite{su_saprot_2024,Rives2021,brandes_proteinbert_2022}. This unsupervised training teaches a transformer to encode protein sequences into an embedding with high performance on a number of downstream tasks such as structure, function, and mutation prediction\cite{liu2022protein,bernardes2013review,fang2022helixfold}. To demonstrate the utility of the PLM within SPRINT, we replace the PLM embedding with a one-hot encoding of the protein sequence, keeping the learned aggregation layer the same. We show in Table \ref{tab:PLMablate} that one-hot encoding drops the DTI prediction performance in comparison to PLMs like ProtBert and SaProt. SPRINT demonstrates yet another use-case where the evolutionary information implicitly understood by the PLM outperforms a one-hot encoding baseline.

\begin{table}[th]
    \caption{Ablating PLM and effect on DTI classification performance (AUPR), SPRINT-sm}
    \label{tab:PLMablate}
    \centering
    \begin{tabular}{cccc}
    \toprule
         Protein Model & BindingDB & BioSNAP & DAVIS \\
    \midrule
        one-hot & 0.546 $\pm$ 0.003 & 0.716 $\pm$ 0.002 & 0.210 $\pm$ 0.003 \\
        ProtBert & 0.672 $\pm$ 0.003 & 0.904 $\pm$ 0.005 & 0.493 $\pm$ 0.014 \\
        \textbf{SaProt} & \textbf{0.718} $\pm$ 0.0004 & \textbf{0.936} $\pm$ 0.001 & \textbf{0.507} $\pm$ 0.005 \\
    \bottomrule 
    \end{tabular}
    
\end{table}

The MERGED dataset splits were determined by clustering protein sequences using MMSeqs2 \citep{steinegger2017mmseqs2} at 80\% coverage threshold and 70\% sequence identity, meaning two sequences appear in the same cluster if at least 80\% of residues are aligned with at least 70\% identity. Clusters were then assigned to splits by size, with smaller clusters preferentially assigned to test and validation sets until each contained approximately 10\% of the total number of unique proteins. The remaining sequences were assigned to training. Drug-target interactions were then partitioned according to their protein assignments. The final training, validation, and test sets contained 79.5\%, 10.3\%, and 10.2\% of total interactions, respectively.

\section{Virtual Screening Experiments}

\begin{figure}[H]
    \centering
    \begin{minipage}{\textwidth}
        \centering
        \textbf{PPARG}\\[0.2em]
        \includegraphics[width=\textwidth]{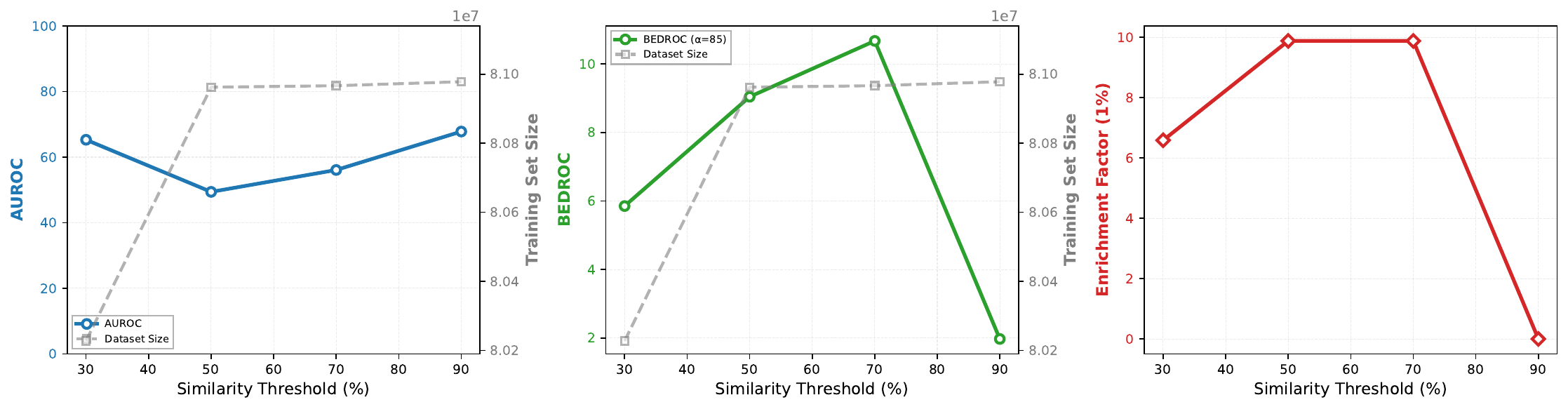}
    \end{minipage}
    \par\vspace{0.5em}
    \begin{minipage}{\textwidth}
        \centering
        \textbf{TP53}\\[0.2em]
        \includegraphics[width=\textwidth]{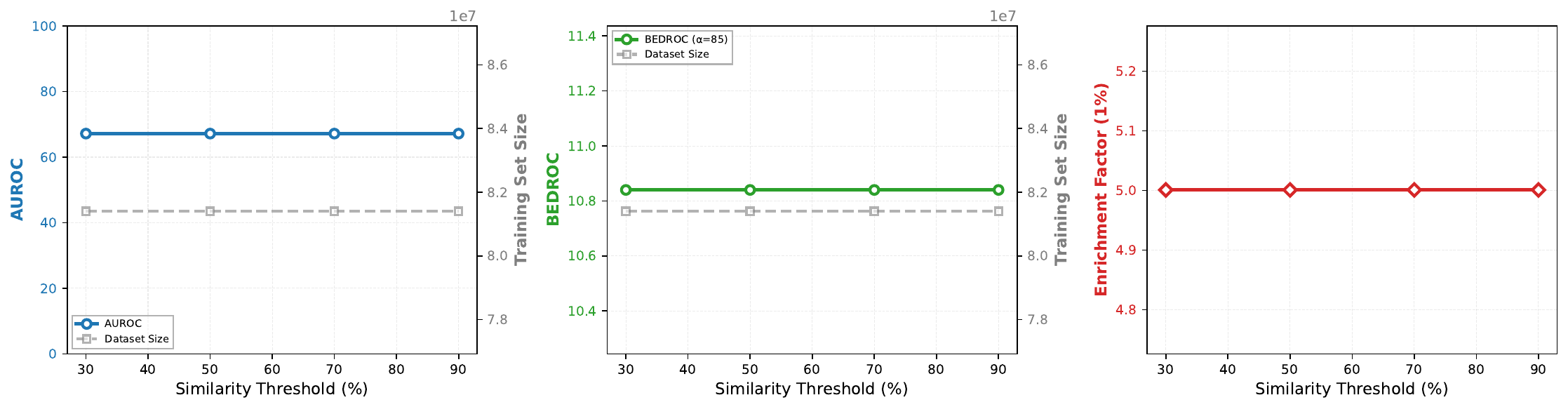}
    \end{minipage}
    \par\vspace{0.5em}
    \begin{minipage}{\textwidth}
        \centering
        \textbf{VDR}\\[0.2em]
        \includegraphics[width=\textwidth]{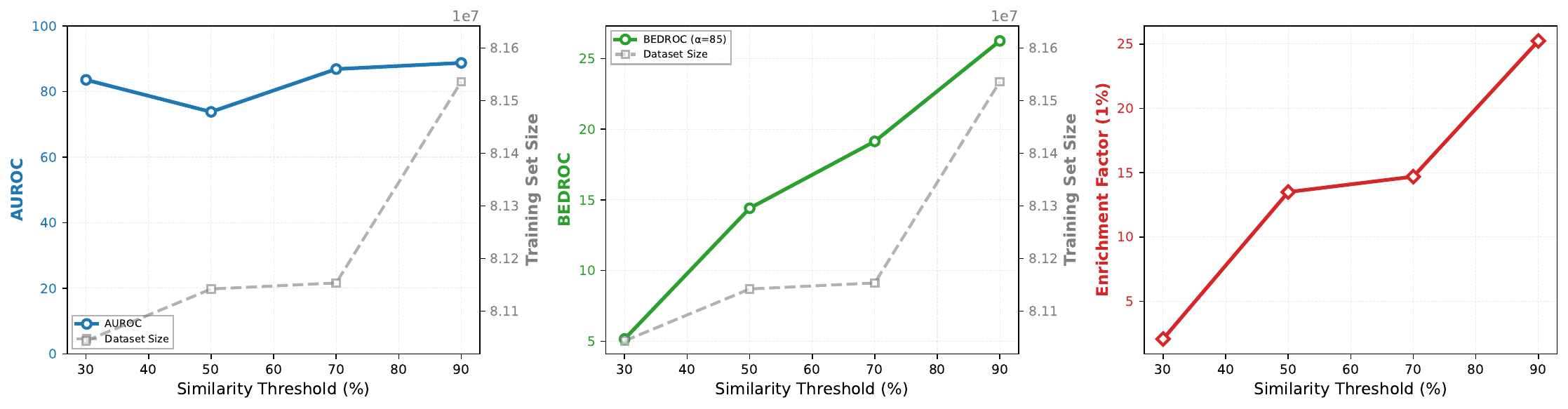}
    \end{minipage}
    \par\vspace{0.5em}
    \caption{LIT-PCBA performance with changing similarity threshold}
    \label{fig:litpcba}
\end{figure}

To assess whether SPRINT generalizes to proteins with lower sequence identity to the training proteins, we evaluated each of the 15 LIT-PCBA targets individually at identity thresholds of 90\%, 70\%, 50\%, and 30\%. For each target, we combined the MERGED training, validation, and test sets and used MMseqs2 to identify training proteins matching any reference sequence associated with that target at or above the specified threshold. We applied these target-specific exclusions and trained a separate model for each target--threshold combination. Compounds were ranked by cosine similarity between ligand and target embeddings, taking the maximum similarity across reference sequences when multiple sequences were available. Performance was assessed using AUROC, BEDROC, and enrichment factors at 1\%. Averaged across targets, performance did not decrease monotonically with stricter filtering (Figure~\ref{fig:litpcba}). When the sequence-identity threshold was reduced from 90\% to 30\%, mean AUROC increased from 68.23\% to 69.75\%, while mean BEDROC decreased from 9.38\% to 8.60\% and mean EF at 1\% decreased from 8.18 to 7.49. These results suggest that SPRINT retains useful virtual-screening performance when closely related training proteins are excluded, although the effect varies across targets.

\begin{figure}[H]
    \centering
    \begin{minipage}{\textwidth}
        \centering
        \textbf{ADRB2}\\[0.2em]
        \includegraphics[width=\textwidth]{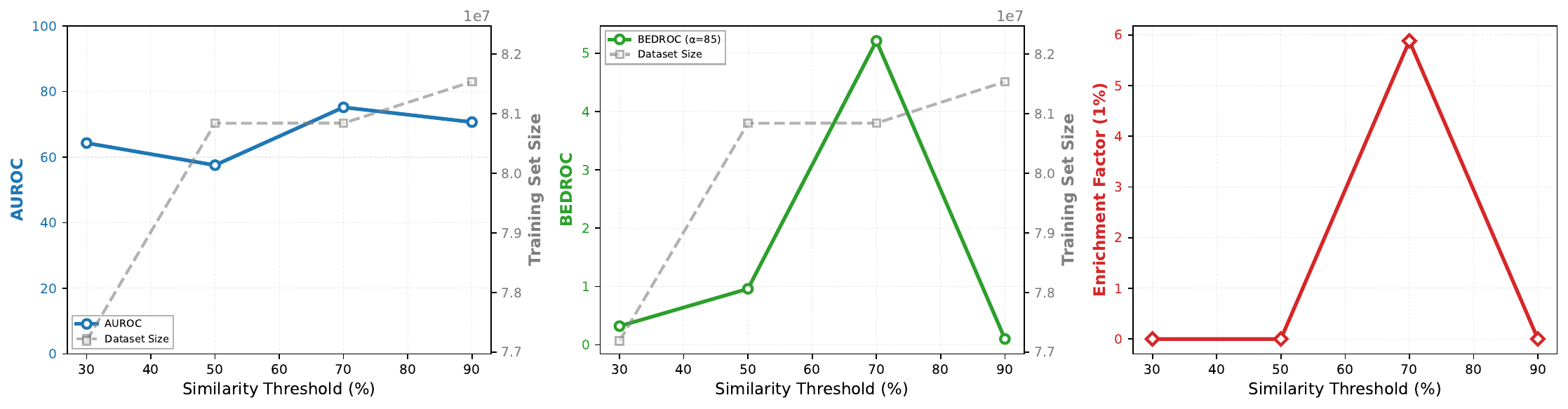}
    \end{minipage}
    \par\vspace{0.5em}
    \begin{minipage}{\textwidth}
        \centering
        \textbf{ALDH1}\\[0.2em]
        \includegraphics[width=\textwidth]{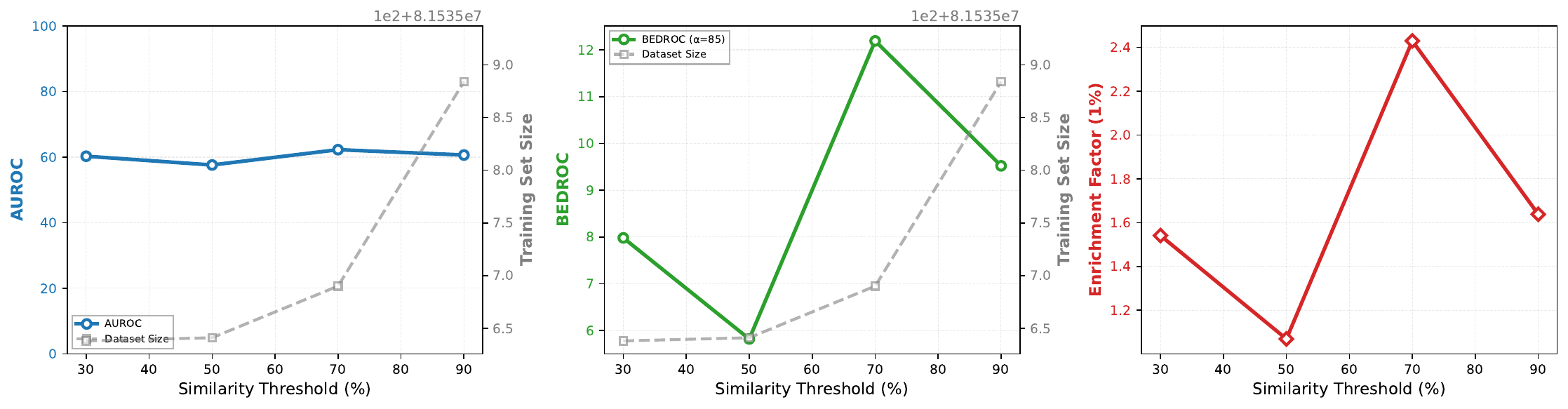}
    \end{minipage}
    \par\vspace{0.5em}
    \begin{minipage}{\textwidth}
        \centering
        \textbf{ESR1_ago}\\[0.2em]
        \includegraphics[width=\textwidth]{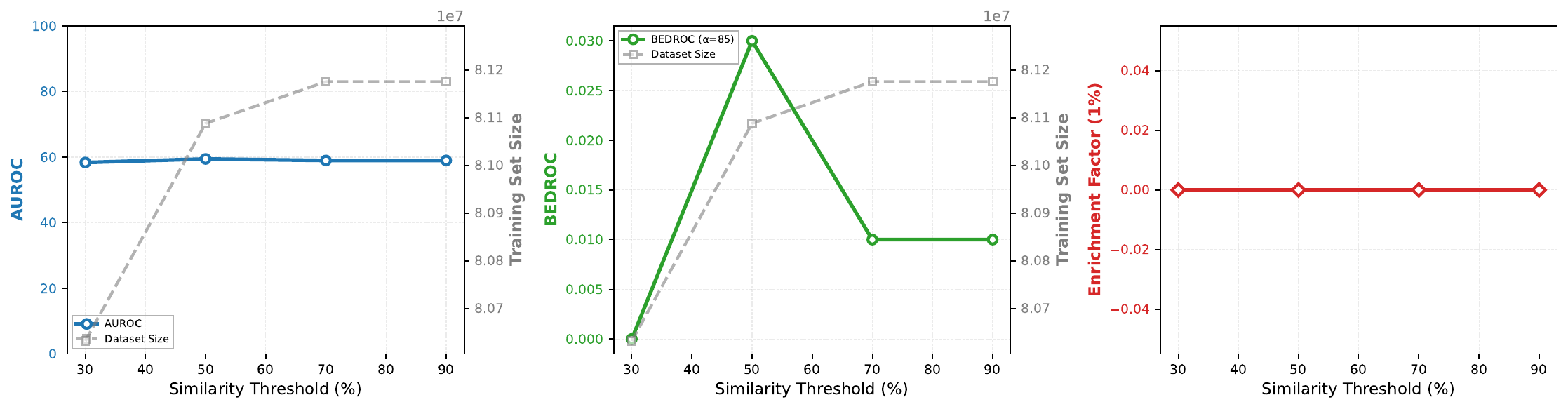}
    \end{minipage}
    \par\vspace{0.5em}
    \begin{minipage}{\textwidth}
        \centering
        \textbf{ESR1_ant}\\[0.2em]
        \includegraphics[width=\textwidth]{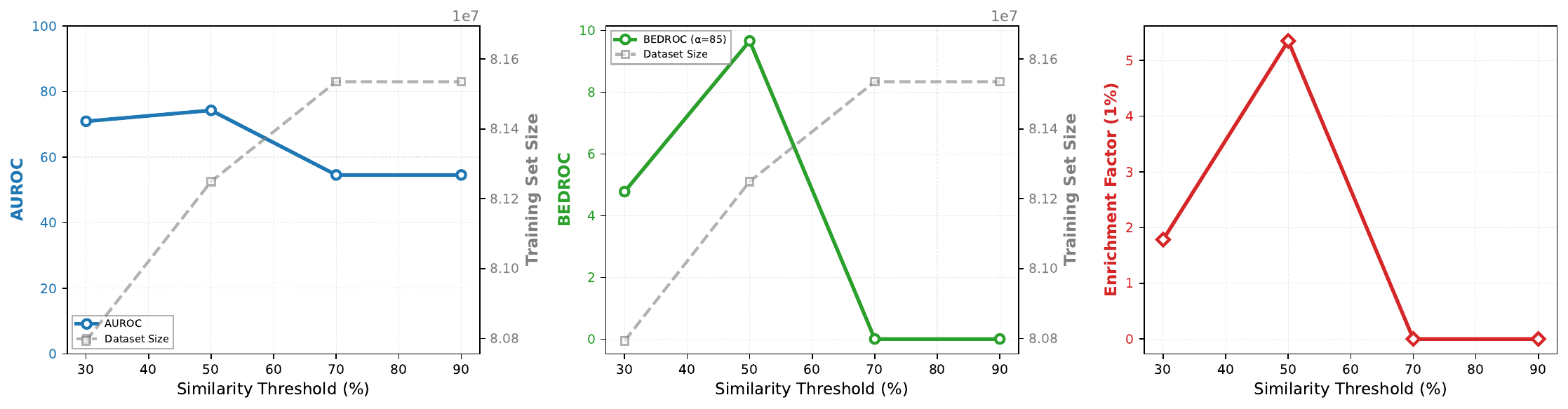}
    \end{minipage}
    \par\vspace{0.5em}
\end{figure}

\begin{figure}[H]
    \centering
    \begin{minipage}{\textwidth}
        \centering
        \textbf{FEN1}\\[0.2em]
        \includegraphics[width=\textwidth]{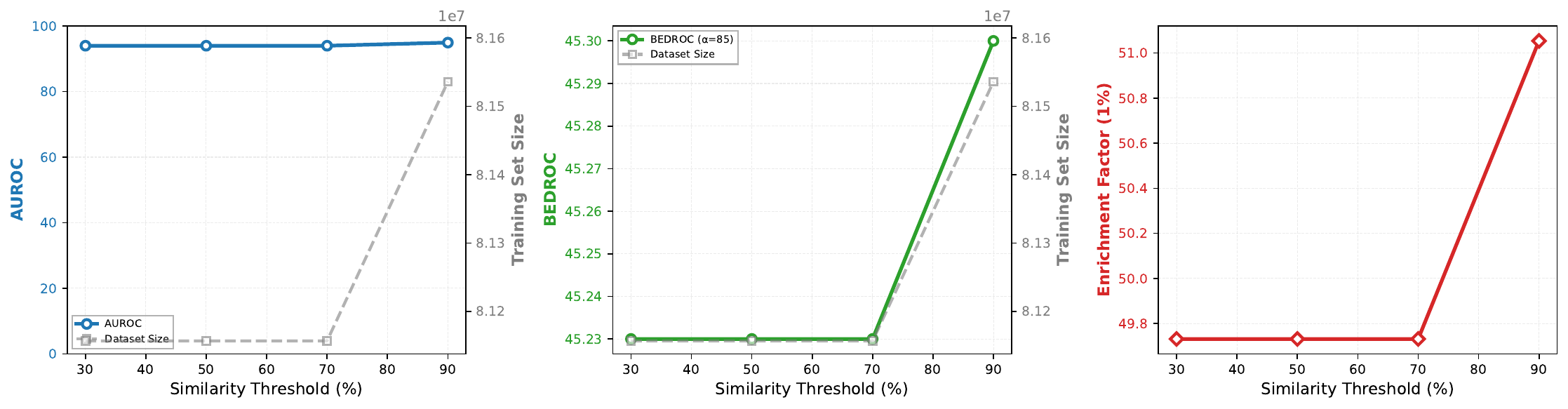}
    \end{minipage}
    \par\vspace{0.5em}
    \begin{minipage}{\textwidth}
        \centering
        \textbf{GBA}\\[0.2em]
        \includegraphics[width=\textwidth]{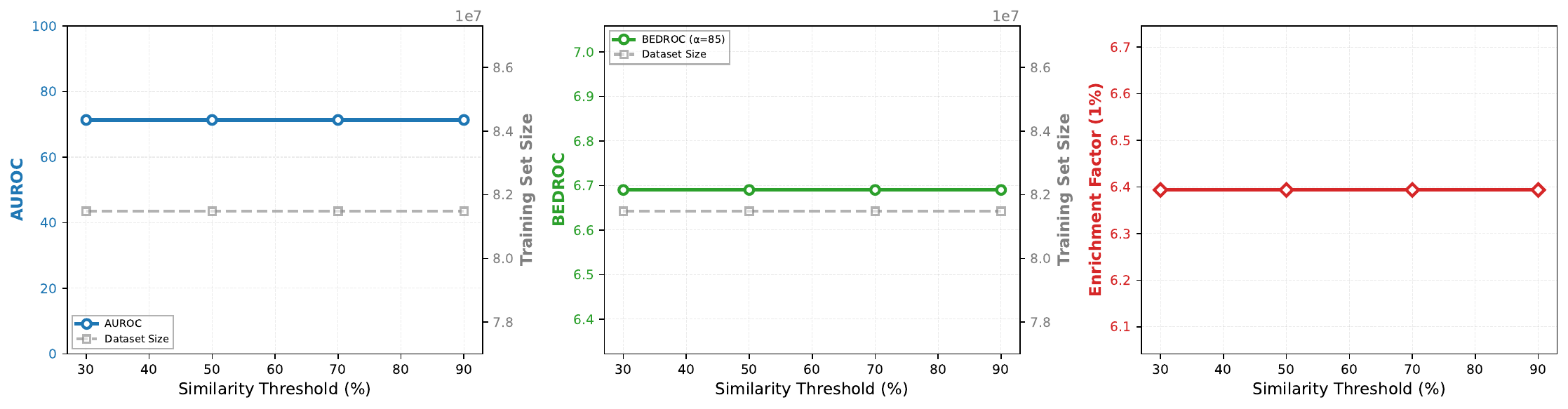}
    \end{minipage}
    \par\vspace{0.5em}
    \begin{minipage}{\textwidth}
        \centering
        \textbf{IDH1}\\[0.2em]
        \includegraphics[width=\textwidth]{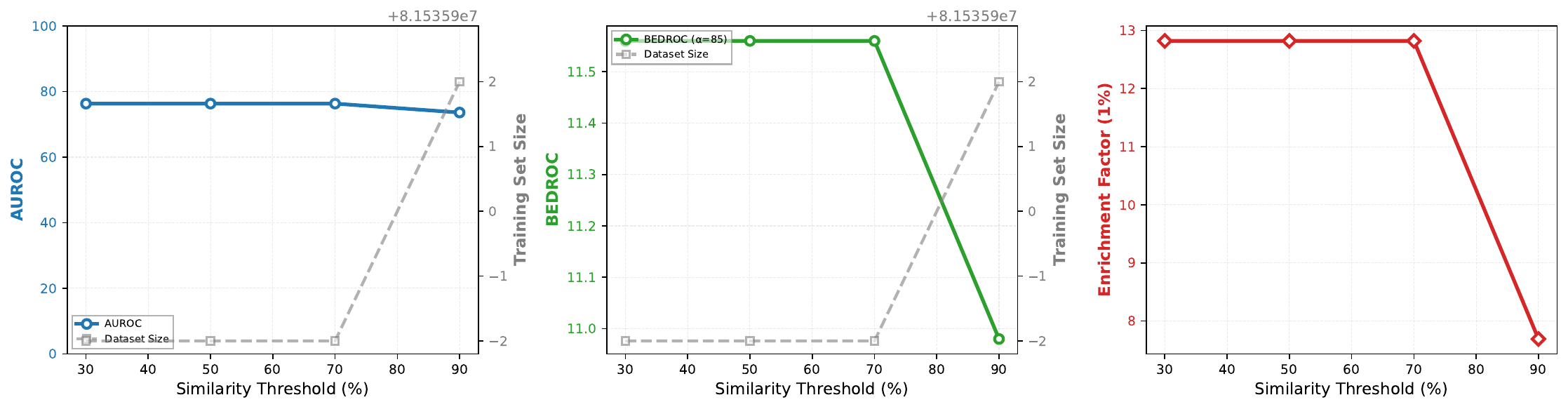}
    \end{minipage}
    \par\vspace{0.5em}
    \begin{minipage}{\textwidth}
        \centering
        \textbf{KAT2A}\\[0.2em]
        \includegraphics[width=\textwidth]{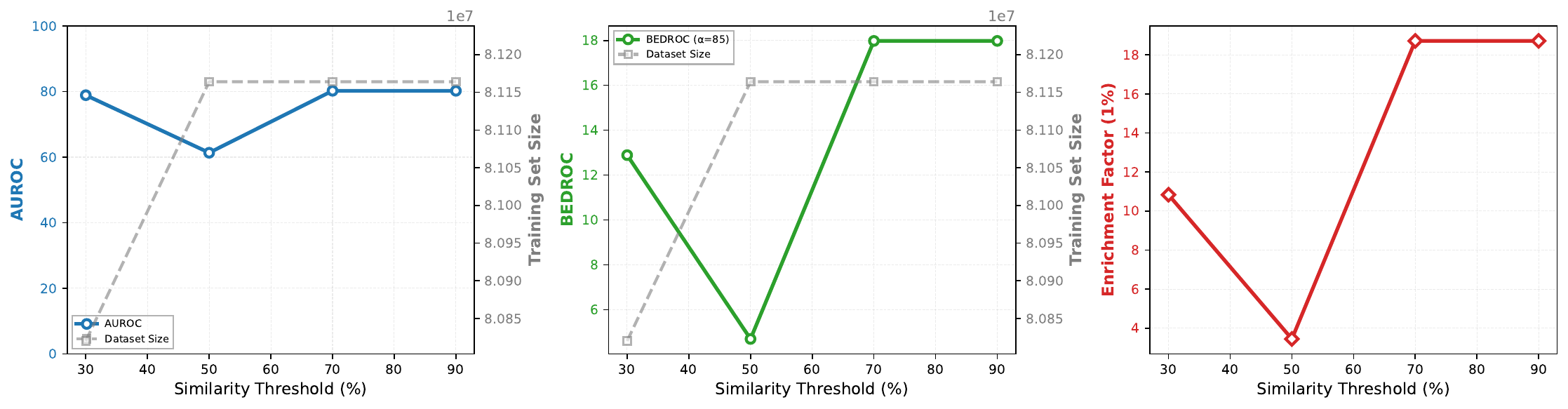}
    \end{minipage}
    \par\vspace{0.5em}
\end{figure}

\begin{figure}[H]
    \centering
    \begin{minipage}{\textwidth}
        \centering
        \textbf{MAPK1}\\[0.2em]
        \includegraphics[width=\textwidth]{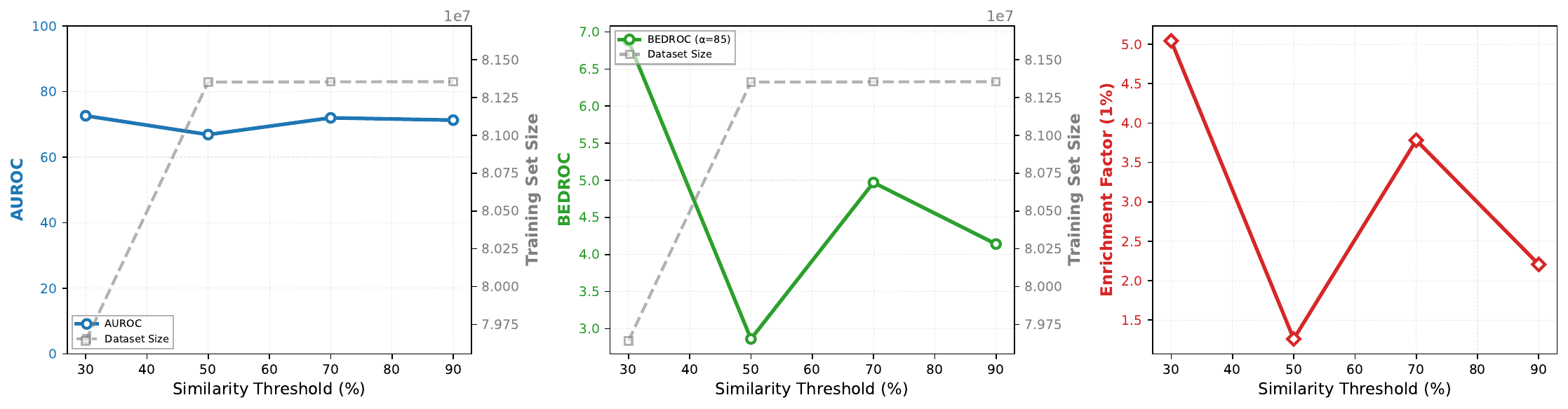}
    \end{minipage}
    \par\vspace{0.5em}
    \begin{minipage}{\textwidth}
        \centering
        \textbf{MTORC1}\\[0.2em]
        \includegraphics[width=\textwidth]{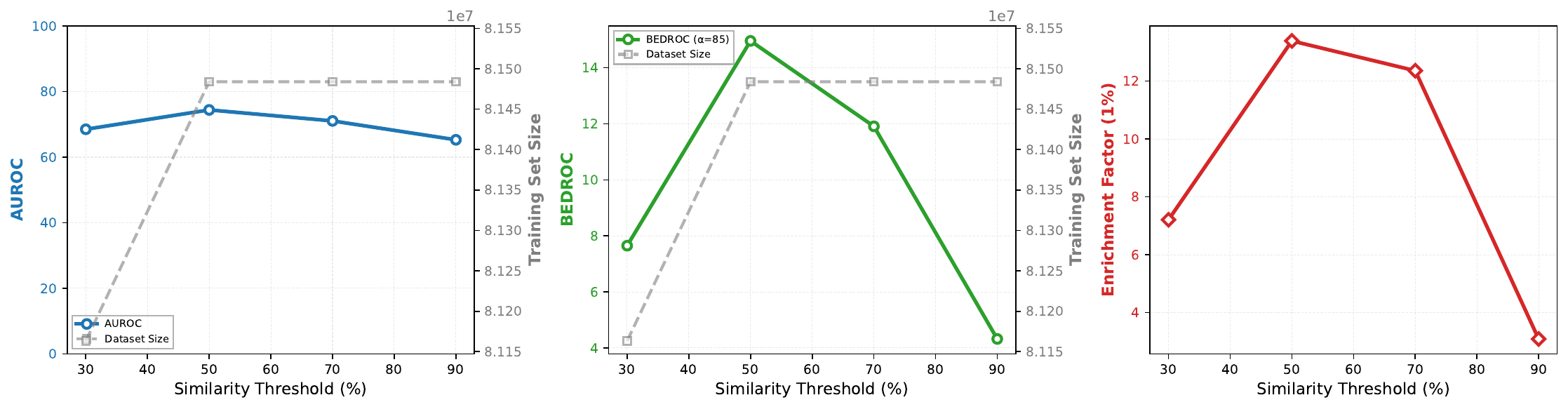}
    \end{minipage}
    \par\vspace{0.5em}
    \begin{minipage}{\textwidth}
        \centering
        \textbf{OPRK1}\\[0.2em]
        \includegraphics[width=\textwidth]{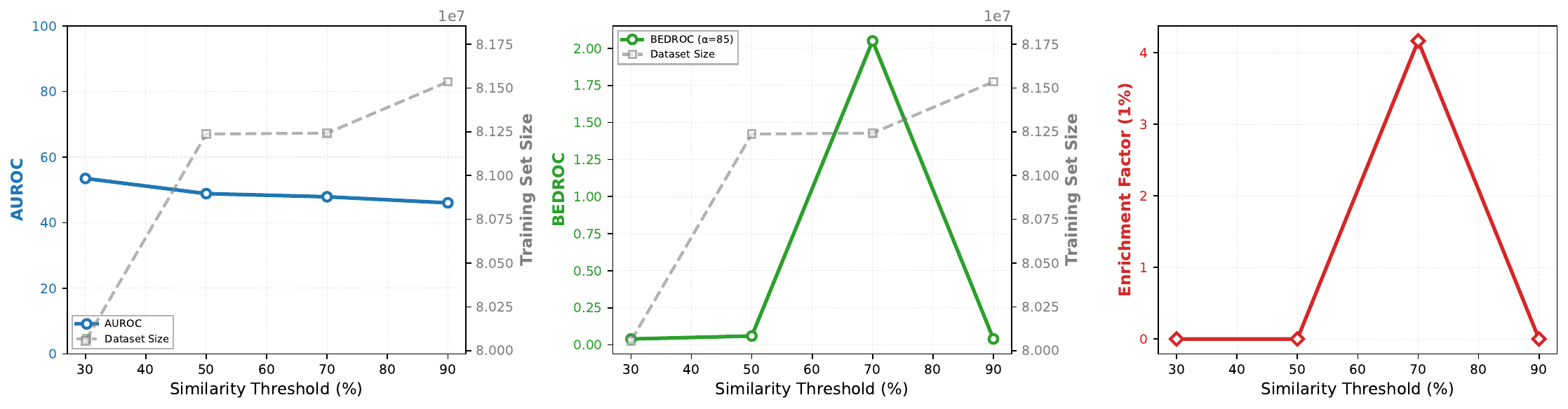}
    \end{minipage}
    \par\vspace{0.5em}
    \begin{minipage}{\textwidth}
        \centering
        \textbf{PKM2}\\[0.2em]
        \includegraphics[width=\textwidth]{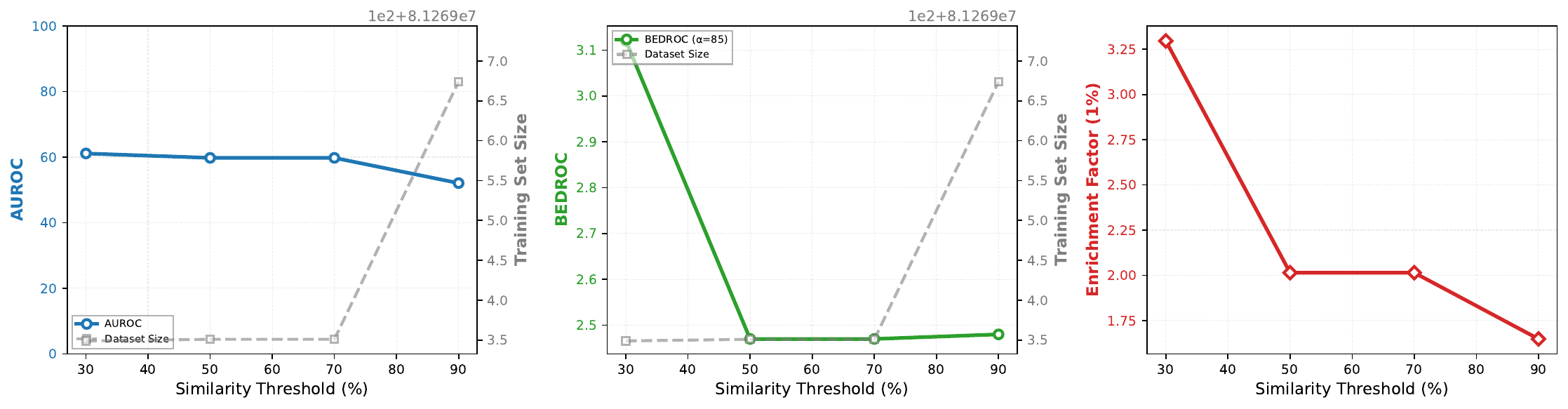}
    \end{minipage}
    \par\vspace{0.5em}
\end{figure}

\begin{table}[h]
\caption{LIT-PCBA evaluation ablation of negative sampling. `1:1' indicates equal sampling of positive and negative examples during training and `3:1' indicates the preferred model training with 3 negatives sampled for every positive example.}
\label{tab:noneg_sample}
\bigskip
\centering
\begin{tabular}{cccccc}
\toprule
Model & AUROC & BEDROC ($\alpha = 85$) & EF (0.5\%) & EF (1\%) & EF (5\%) \\
\midrule
SPRINT-ProtBert 1:1 & 71.53 & 7.78 & 6.81 & 5.87 & 3.86 \\
SPRINT 1:1 & 72.71 & 10.16 & 10.31 & 8.86 & 4.73 \\

\midrule
SPRINT-ProtBert 3:1 & \textbf{73.4} & 11.9 & 11.68 & 10.19 & 5.27 \\
SPRINT 3:1 & \textbf{73.4} & \textbf{12.3} & \textbf{15.90} & \textbf{10.78} & \textbf{5.29} \\
\bottomrule
\end{tabular}
\end{table}
The effect of negative sampling was evaluated by training SPRINT and SPRINT-ProtBert with negative-to-positive sampling ratios of 1:1 and 3:1, corresponding to one or three negative examples per positive example. The resulting models were evaluated on LIT-PCBA using AUROC, BEDROC, and EF at 1\%. Increasing the sampling ratio from 1:1 to 3:1 improved all three metrics for both models (Table ~\ref{tab:noneg_sample}). EF at 1\% increased from 8.86 to 10.78 for SPRINT and from 5.87 to 10.19 for SPRINT-ProtBert. The 3:1 ratio was therefore used as the preferred negative sampling configuration.

\begin{table}[th]
    \caption{Enrichment factor virtual screening performance on RandomDecoy and TrueDecoy datasets \cite{gu2025benchmarking}.}
    \label{tab:DecoyVS}
    \centering
    \small
    \begin{tabular}{lcccccccccc}
         \toprule
         & \multicolumn{3}{c}{TrueDecoy} & \multicolumn{3}{c}{RandomDecoy} \\
        \cmidrule(r){2-4} \cmidrule(r){5-7}
         & 0.5\% & 1\% & 5\% & 0.5\% & 1\% & 5\% \\
        \midrule
        SPRINT & 1.88 & 1.94 & 1.81 & 45.8 & 33.1 & 11.0 & & &\\
        KarmaDock & 4.45 & 4.28 & 2.93 & 51.7 & 39.9 & 11.8 &\\
        CarsiDock & 5.11 & 4.85 & 3.17 & 59.3 & 43.5 & 12.6 &\\
        Glide\_RTMScore & 6.86 & 5.69 & 3.30 & 51.0 & 34.4 & 9.9 &\\
        Glide\_EquiScore & 3.25 & 3.13 & 2.06 & 27.9 & 19.8 & 7.3 & &\\
    \end{tabular}    
\end{table}

Virtual-screening performance beyond LIT-PCBA was evaluated using the RandomDecoy and TrueDecoy datasets \cite{gu2025benchmarking}. SPRINT was evaluated using the same embedding-based procedure as for LIT-PCBA, with enrichment factors averaged across targets. Results for the comparator methods were taken directly from the paper \cite{gu2025benchmarking}, while SPRINT results were computed in this study (Table~\ref{tab:DecoyVS}). Mean EF at 1\% was 33.1 on RandomDecoy and 1.94 on TrueDecoy, indicating enrichment above random selection on both datasets, with stronger enrichment on RandomDecoy.

\section{High scoring molecules screened with SPRINT for NSP13 Helicase}
\begin{figure}[ht]
    \centering
    \includegraphics[width=\linewidth]{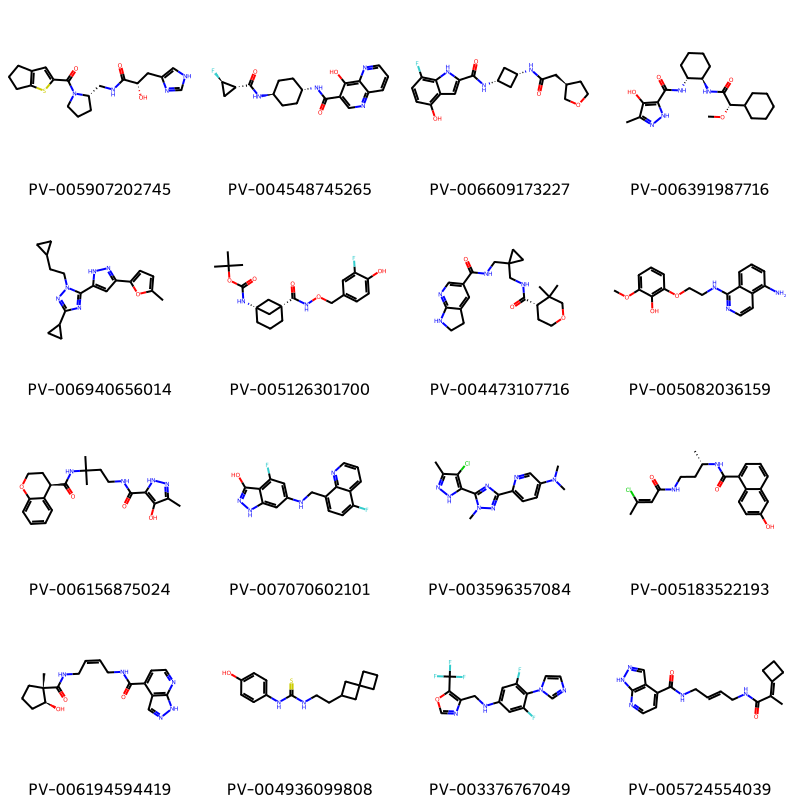}
    \caption{2D depiction of the high-scoring (CNN VS $>$ 6) molecules found by SPRINT for NSP13 Helicase. Images generated using RDKit\cite{rdkit}.}
    \label{fig:SPRINT_molecules}
\end{figure}

We show the 2D depiction of the highest scoring molecules, according to \textsc{Gnina}'s CNN VS, from SPRINT's screen of Enamine REAL in Figure~\ref{fig:SPRINT_molecules}. We see that the molecules represent diverse scaffolds.

\section{Investigating the learned aggregation layer}
\label{sec:Attn_interrogation}
We investigate the attention pattern of our learned aggregation layer and compute the relative weighting of binding and non-binding residues. For this analysis, we use the intersection of the PDBbind refined v.2019 \citep{liu2017forging} dataset and the dataset created by \citep{gazizov2023af2bind}. The intersection of these datasets provides 109 single-chain, high-quality protein-ligand binding structures with annotated binding sites. Following the same protocol as \citep{gazizov2023af2bind}, we determine binding residues based on a maximum heavy-atom distance of 5 {\AA} between the residue and the ligand.

\begin{figure}[ht]
    \centering
    \includegraphics[width=\textwidth]{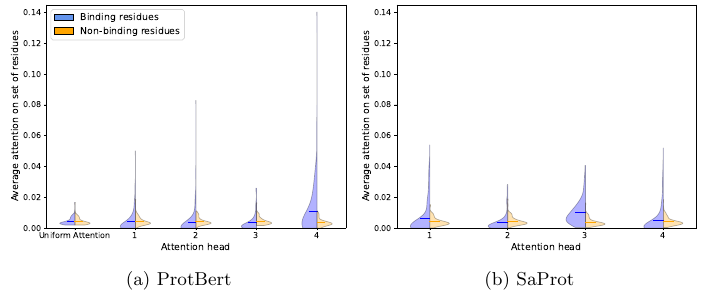}
    \caption{Comparing the average attention weight of binding and non-binding residues on our set of 109 single-chain protein-ligand binding structures after training on the MERGED Dataset (Methodology detailed in Appendix~\ref{sec:Attn_interrogation}). We visualize the Protbert and SaProt models trained with equal positive and negative sampling. The horizontal line indicates the average across the proteins. Visualizations of the ProtBert and SaProt models trained with increased negative sampling are in Figure~\ref{fig:bs_attn_fullfig}).}
    \label{fig:attention-heads}
\end{figure}

\begin{figure}
    \centering
    \includegraphics[width=\textwidth]{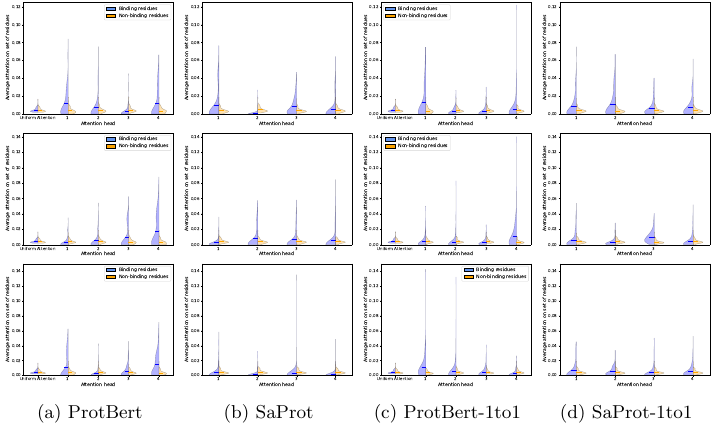}
    \caption{Comparing the average attention weight of binding and non-binding residues on our set of 109 single-chain protein-ligand binding structures after training on the MERGED Dataset. The horizontal line indicates the average across the proteins. Each row is a different random seed and each column is a different PLM or different training regime, where `1to1' indicates that a 1:1 positive to negative sampling ratio was used during training.}
    \label{fig:bs_attn_fullfig}
\end{figure}

We first analyzed the attention patterns of each head in the learned aggregation layer to determine if any of the heads were selective for binding residues. We calculate the attention scores for each residue in the protein and compute the mean attention value for the binding residues and the non-binding residues. Figure~\ref{fig:attention-heads} shows the average weight of the binding residues and non-binding residues across the dataset for the ProtBert and SaProt models trained with equal positive and negative sampling. We find the models trained with equal positive and negative sampling have more interpretable attention maps despite a decreased performance on the LIT-PCBA benchmark compared to models trained with more negative samplings. We compare the attention of all the models in Figure~\ref{fig:bs_attn_fullfig}, visualizing three otherwise identical models trained with different initial random seeds. Interestingly, the PLM with the worst performance on the LIT-PCBA benchmark, ProtBert, shows the most attention to binding site residues relative to its attention to non-binding site residues. The structure-aware PLM, SaProt, has two seeds that attend to binding residues more than non-binding residues across most of the attention heads and one seed that pays very little attention to the binding residues. The SaProt seed that has the least attention for binding residues as compared to non-binding residues performs the best on the LIT-PCBA benchmark. Across all PLMs, there is a large variance in the attention to binding residues as the models initial random seed is changed.

\section{Structural visualizations}
We visualize the attention patterns of the attention pooling layer on the protein-ligand bound structure of several PDB IDs. We compare the attention patterns of SPRINT models with ProtBert and SaProt trained on the MERGED dataset. We see across these diverse proteins and ligands that on average, the SaProt model attends to residues closer to the ligand, while the ProtBert models often attend to residues far from the binding site. 
\label{sec:addnl_structs}
\begin{figure}
    \centering
    \includegraphics[width=\textwidth]{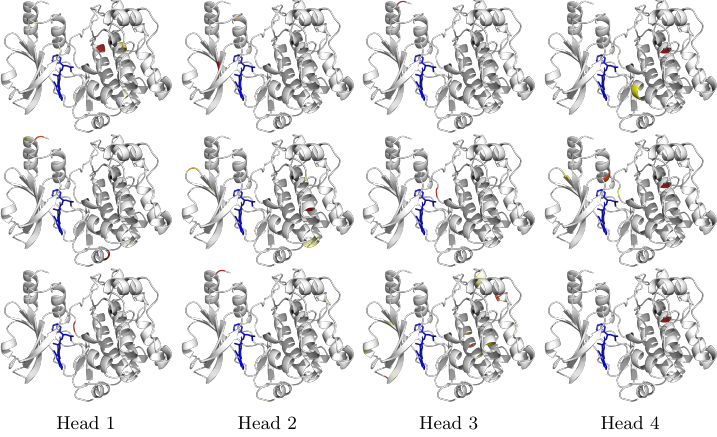}
    \caption{Analyzing the attention of the ProtBert model on PDB ID 2X4Z. Each row is the ProtBert model trained with different seed. Each column is a different attention head. Gradient from white to red indicates the attention weight, where white is no attention and red is max attention for that head. The ligand is shown in blue. Images generated using PyMOL\cite{PyMOL}.}
    \label{fig:struct_attn_protbert_2X4Z}
\end{figure}
\begin{figure}
    \centering
    \includegraphics[width=\textwidth]{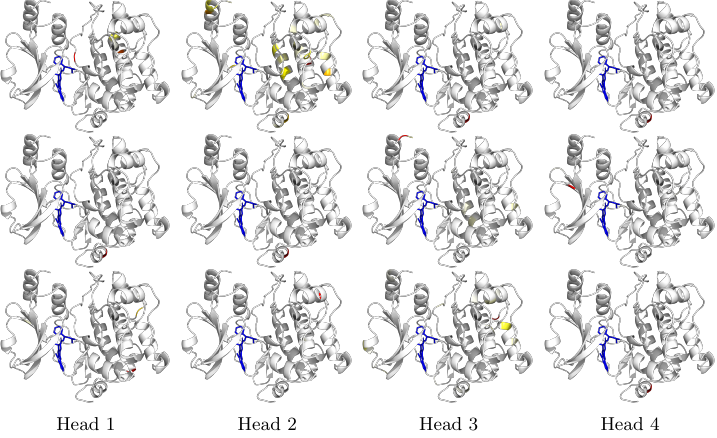}
    \caption{Analyzing the attention of the ProtBert-1to1 model (trained with 1:1 positive to negative ratio) on PDB ID 2X4Z. Each row is the ProtBert model trained with different seed. Each column is a different attention head. Gradient from white to red indicates the attention weight, where white is no attention and red is max attention for that head. The ligand is shown in blue. Images generated using PyMOL\cite{PyMOL}.}
    \label{fig:struct_attn_protbert_one_2X4Z}
\end{figure}
\begin{figure}
    \centering
    \includegraphics[width=\textwidth]{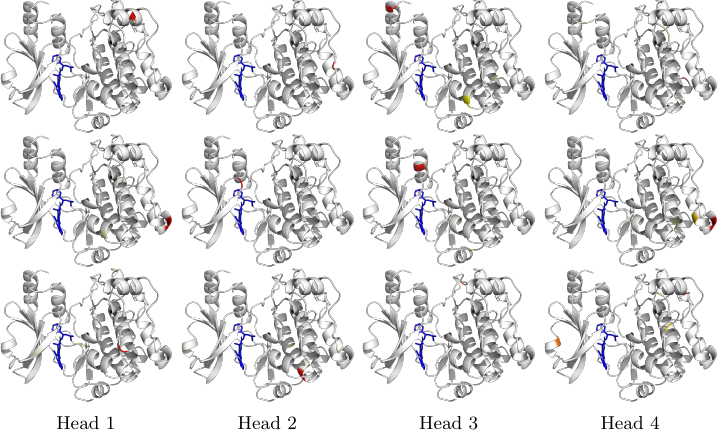}
    \caption{Analyzing the attention of the SaProt model on PDB ID 2X4Z. Each row is the SaProt model trained with different seed. Each column is a different attention head. Gradient from white to red indicates the attention weight, where white is no attention and red is max attention for that head. The ligand is shown in blue. Images generated using PyMOL\cite{PyMOL}.}
    \label{fig:struct_attn_saprot_2X4Z}
\end{figure}
\begin{figure}
    \centering
    \includegraphics[width=\textwidth]{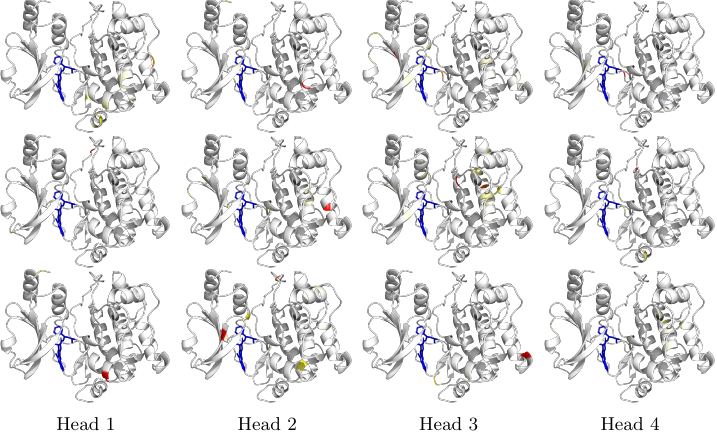}
    \caption{Analyzing the attention of the SaProt-1to1 model (trained with 1:1 positive to negative ratio) on PDB ID 2X4Z. Each row is the SaProt model trained with different seed. Each column is a different attention head. Gradient from white to red indicates the attention weight, where white is no attention and red is max attention for that head. The ligand is shown in blue. Images generated using PyMOL\cite{PyMOL}.}
    \label{fig:struct_attn_saprot_one_2X4Z}
\end{figure}



\end{appendices}
\clearpage

\bibliography{sn-bibliography}

\end{document}